\documentclass{emulateapj}
\pdfoutput=1

\def\alwaysmath#1{\ifmmode{#1}\else{$#1$}\fi}

\def\beq{\begin{equation}}
\def\eeq{\end{equation}}
\def\beqa{\begin{eqnarray}}
\def\eeqa{\end{eqnarray}}
\def\lsim{\lower0.6ex\vbox{\hbox{$ \buildrel{\textstyle <}\over{\sim}\ $}}}
\def\gsim{\lower0.6ex\vbox{\hbox{$ \buildrel{\textstyle >}\over{\sim}\ $}}}

\begin{document}

\title{Tracing Galaxy Formation with Stellar Halos II: \\ 
Relating
Substructure in Phase- and Abundance-Space
to Accretion Histories}
\author{Kathryn V. Johnston \altaffilmark{1} James  S. Bullock\altaffilmark{2}, Sanjib Sharma\altaffilmark{1}, Andreea Font\altaffilmark{3}, Brant E. Robertson\altaffilmark{4,5,6} and  Samuel N. Leitner\altaffilmark{4}}
\altaffiltext{1}{Department of Astronomy, Columbia University, Pupin Physics Laboratory, 550 West 120th Street, New York, NY 10027, USA}
\altaffiltext{2}{Center for Cosmology, Department of Physics \& Astronomy,
        University of California, Irvine, CA 92697, USA; bullock@uci.edu}
\altaffiltext{3}{Institute for Computational Cosmology, University of Durham, South Road, Durham DH1 3LE, United Kingdom}
\altaffiltext{4}{Kavli Institute for Cosmological Physics, Department of Astronomy and Astrophysics, University of Chicago, 933 East 56th Street,
Chicago, IL 60637}
\altaffiltext{5}{Enrico Fermi Institute, 5640 S. Ellis Ave.
Chicago, IL 60637}
\altaffiltext{6}{Spitzer Fellow}

\begin{abstract}
This paper explores the mapping between the observable properties of a stellar halo in phase- and abundance-space and the parent galaxy's accretion history in terms of the characteristic epoch of accretion and mass and orbits of progenitor objects.
The study utilizes a suite of eleven stellar halo models constructed within the context of a standard $\Lambda$CDM cosmology.
The results demonstrate that coordinate-space studies are sensitive to the recent (0-8 Gyears ago) merger histories of galaxies (this timescale corresponds to the last few to tens of percent of mass accretion for a Milky-Way-type galaxy). 
Specifically, the {\it frequency, sky coverage} and {\it fraction of stars} in substructures in the stellar halo as a function of surface brightness are indicators of the importance of recent merging and of the luminosity function of infalling dwarfs. 
The {\it morphology} of features serves as a guide to the orbital distribution of those dwarfs. 
Constraints on the earlier merger history ($>$ 8  Gyears ago) can be gleaned from the abundance patterns in halo stars: within our models, dramatic differences in the dominant epoch of accretion  or luminosity function of progenitor objects leave clear signatures in the [$\alpha$/Fe] and [Fe/H] distributions of the stellar halo --- halos dominated by very early accretion have higher average [$\alpha$/Fe], while those dominated by high luminosity satellites have higher [Fe/H]. 
This intuition can be applied to reconstruct much about the merger histories of nearby galaxies from current and future data sets.


\end{abstract}
\keywords{Galaxy: evolution  ---  Galaxy: formation ---  Galaxy:halo --- Galaxy:
kinematics  and dynamics ---  galaxies:  dwarf --- galaxies: evolution
--- galaxies: formation  --- galaxies: halos  --- galaxies: kinematics
and dynamics --- Local Group --- dark matter}

\section{Introduction}

Phase- and abundance-space substructure in the stellar distribution  around a galaxy is commonly interpreted as a natural outcome of the process of hierarchical structure formation, where large galaxies are built  in part from the accretion of dwarf galaxies.
Numerous previous studies have looked at how to understand individual debris features around galaxies in terms of the properties of the progenitor dwarfs  \citep[e.g.][]{johnston98,helmi99a,johnston01,law05,fardal06,warnick08}.
More generally, we might ask: to what extent can the merger history of a galaxy be reconstructed from its surrounding substructure?; and what could  you learn about the nature of merger histories in our Universe by examining the outskirts of a large sample of galaxies to very low surface brightness?
These questions have yet to be addressed beyond using simple analytic estimates for the expected scalings in tidal debris \citep{bullock01,johnston01}.

Motivation for answering these questions in more depth has been amply provided in the last decade by
observations which have revealed abundant substructure in the spatial distribution around both the Milky Way and Andromeda galaxies, already seen at a few tens of kpc from their centers, and dominant at 100kpc \citep[e.g.][and see \S \ref{current.sec} for more examples]{ivezic00,newberg02,newberg03,ferguson02,majewski03,belokurov06,ibata07}. The ubiquity of such substructure has become apparent because of dramatic increases in the sky-coverage of halo surveys, the number of tracers (and hence surface brightness to which such surveys are sensitive) and the distances from the parent galaxies which have been probed.
In contrast, only slightly more than a decade ago the classical picture of stellar halos around galaxies  was that the stars in them were smoothly distributed in phase-space
--- a picture informed by observations of the RR Lyrae and globular cluster distribution around the Milky Way using samples of a few dozen to hundreds  of objects \citep[e.g.][]{wetterer96}.
In our own Galaxy, even halo stars in the Solar neighborhood have been shown to be clumped once their full
phase-space coordinates are known and their orbital distribution is considered \citep{helmi99b,morrison08}.

Satellite accretion is the preferred explanation for apparent phase-space substructure.
Debris from the destruction of a satellite will phase-mix along the progenitor's orbit over time to fill the full volume of the original orbit in coordinate space \citep{johnston98}, while becoming locally colder at each point in velocity-space  \citep[in order to satisfy Louiville's theorem --- see][for a rigorous description]{helmi99a}. The small range in orbital angular momenta and energy is largely conserved during any gradual evolution of the parent galaxy potential, although the average values can evolve \citep{penarrubia06}. 
\citep[See also][for a discussion in a more violent context.]{knebe05,warnick08} 
Putting these results for individual accretion events together leads to a spectrum of lumps in phase-space, as has been demonstrated in composite studies of halo distributions \citep{helmi00,bullock01,johnston01,helmi03,bullock05,delucia08}. 

Similarly, substructure is now becoming apparent in metallicity distributions and abundance patterns of halo stars. So far, these studies have relied on using coordinates in both phase- and abundance-space to identify these structures \citep[e.g.][]{majewski96,navarro04,helmi06,ibata07,gilbert08}. However, abundance space has the advantage over phase-space of preserving a complete memory of a star's initial conditions which cannot be destroyed by subsequent mixing or scattering. In principle,``chemically tagging'' stars could lead to associations that are not apparent in phase-space alone \citep{blandhawthorn03}. Indeed, it has already been demonstrated that the distinct merging time and mass scales of stellar halo,  satellite and dominant substructure progenitors  means that stars in each of these systems should have distinct patterns in $\alpha$-element/metallicity space \citep{robertson05,font06a,font06b,font08}, which implies that distinctions between lower-contrast systems could be possible in higher dimensions. These results also suggest that abundance distributions in a stellar halo should reflect its accretion history \citep[as has already been shown for metallicity distributions:][]{renda05,font06b}.

This paper takes a first step towards addressing how to broadly  relate substructures apparent today to a galaxy's past by exploring how the observed properties of eleven stellar halo models built entirely from accretion events within the context of a $\Lambda$CDM universe reflect their accretion histories.
Further work will build on the intuition developed here to define statistical measures of substructure that are tuned to be sensitive to the epoch of accretion and mass and orbit type of progenitor satellites.
The modeling methods are described in \S2. 
In \S 3 the 1515 simulated accretion events from all eleven model halos are first analyzed individually in order to characterize how the {\it observed} properties of debris (i.e. spatial and velocity scales, morphology, surface brightness and abundance patterns) are related to the {\it intrinsic} properties of the progenitor satellite (i.e. accretion time, luminosity and orbit). 
The intuition developed in \S 3 is then applied in  \S 4 to understanding how the {\it observed} properties of stellar halos (i.e. frequency, morphology, surface brightness and stellar populations of non-uniform features) arise from their unique formation histories.  
These ideas are illustrated by contrasting our eleven ``standard'' (i.e. built within a $\Lambda$CDM context) halos with four model halos built from accretion histories that have been artificially constrained to be dominated by ancient/recent and  high/low luminosity events, as well as two more built from events predominantly on radial/circular orbits.
The results are reviewed in \S5, and subsequently applied to ``observations'' of our model halos  in order to demonstrate in principle how such data might be interpreted.
In \S6 we discuss applications of these ideas to both current and future data sets.
We summarize our conclusions in \S7.

\section{Modeling Methods}

Our models were constructed to provide high-resolution examples of what stellar halos around Milky-Way-type galaxies built from accretion within the context of a $\Lambda$CDM universe might look like. A concise description is included below. \citep[For full details of the method and tests of our models see][]{bullock05,robertson05,font06a}

The modeling method can be broadly split into three phases:
\begin{description}
\item{Phase I: {\it Simulating the dark matter dynamical evolution} ---}
The time and mass of each accretion event throughout a galaxy's history was first generated at random from an extended Press-Schechter merger tree \citep{somerville99}.
The events were assigned orbital eccentricities and binding energies drawn from orbital distributions of  satellites observed in fully self-consistent cosmological simulations of structure formation.
Individual, high-resolution N-body simulations were then run to track the evolution of the dark matter in the infalling objects during each of these events, with the parent galaxy represented by a time-dependent, analytic potential consisting of bulge, disk and (spherical) dark matter halo components.
\item{Phase II: {\it Describing the baryonic evolution and final luminous structure} ---  }
Star formation histories were assigned to each accreted object via a simple prescription.
The parameters of the prescription were normalized so that the final stellar and gas content of satellites infalling today were similar to those of the Local Group dwarfs. 
The stars thus generated were ``painted'' on to the dark matter particles so as to reproduce the trends in structural properties of Local Group dwarfs.
\item{Phase III: {\it Following the chemical enrichment history} ---}
The chemical enrichment associated with the star formation within each object was followed using a leaky-accreting-box chemical enrichment code \citep{robertson05}. The parameters of the chemical enrichment model were tuned to reproduce the mass-metallicity relation of Local Group dwarfs and the low alpha-element ratios seen in Milky Way satellites.
\end{description}

With all free parameters now fixed, the resulting stellar halos were found to have
total stellar content ($\sim 10^9 L_\odot$) and density profile comparable to the Milky Way's halo \citep{bullock05} and similar abundance patterns to stars in the Milky Way's halo \citep{font06a}. They also contained satellite systems with roughly the same number of high-surface brightness companions as the Milky Way \citep[neglecting the recent and lower surface-brightness discoveries by][and others]{willman05a,willman05b,belokurov06}, and whose members had  similar structural parameters to the Milky Way's satellites.
Note that more recent work, using semi-analytic models in combination with fully self-consistent cosmological simulations of the formation of Milky-Way-type galaxies has found very similar results \citep{delucia08}

\section{Results I: characteristics of debris from individual accretion events}
\label{resultsi.sec}

In this section we consider each accretion event individually in order to characterize how the spatial and velocity scales, morphology, surface brightness and abundance patterns of debris depend on the accretion time, luminosity and orbit of the progenitor satellite. We first discuss debris from unbound systems before going on to look at debris associated with still-bound systems (in \S 3.5)

\subsection{Morphology}
\label{morph.sec}

\begin{figure*}
\epsscale{1.0}
\plotone{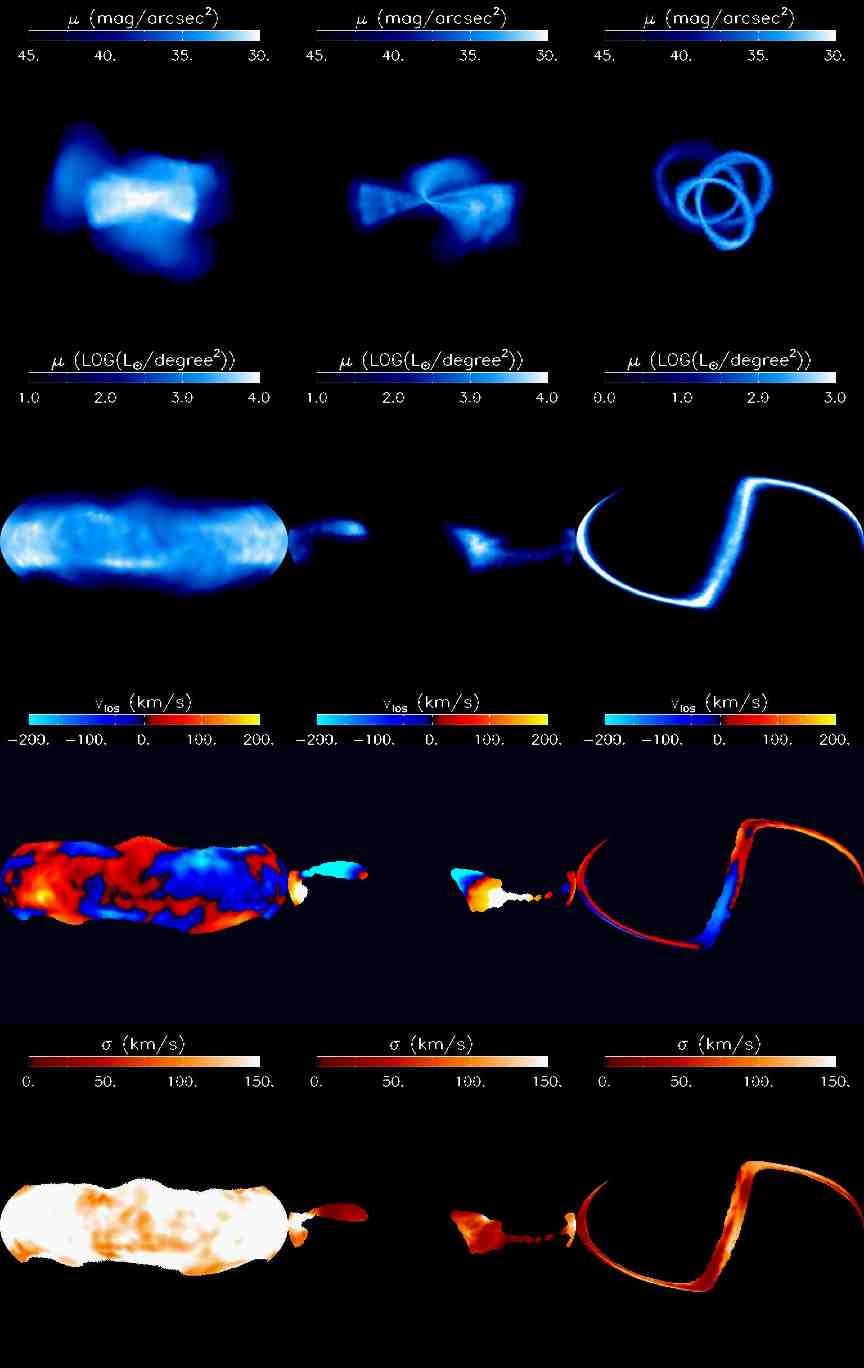}
\caption{\label{morph135.fig}
Plots to illustrate the general characteristics of mixed (left hand panels), shell/plumes/clouds (middle panels) and great circle (right hand panels) morphologies. Top panels show external views in surface brightness (with boxsize scaled arbitrarily). Lower panels show all sky Aitoff projections of surface density, mean line-of-sight velocity and velocity dispersion.}
\end{figure*}

\begin{figure*}
\epsscale{1.0}
\plotone{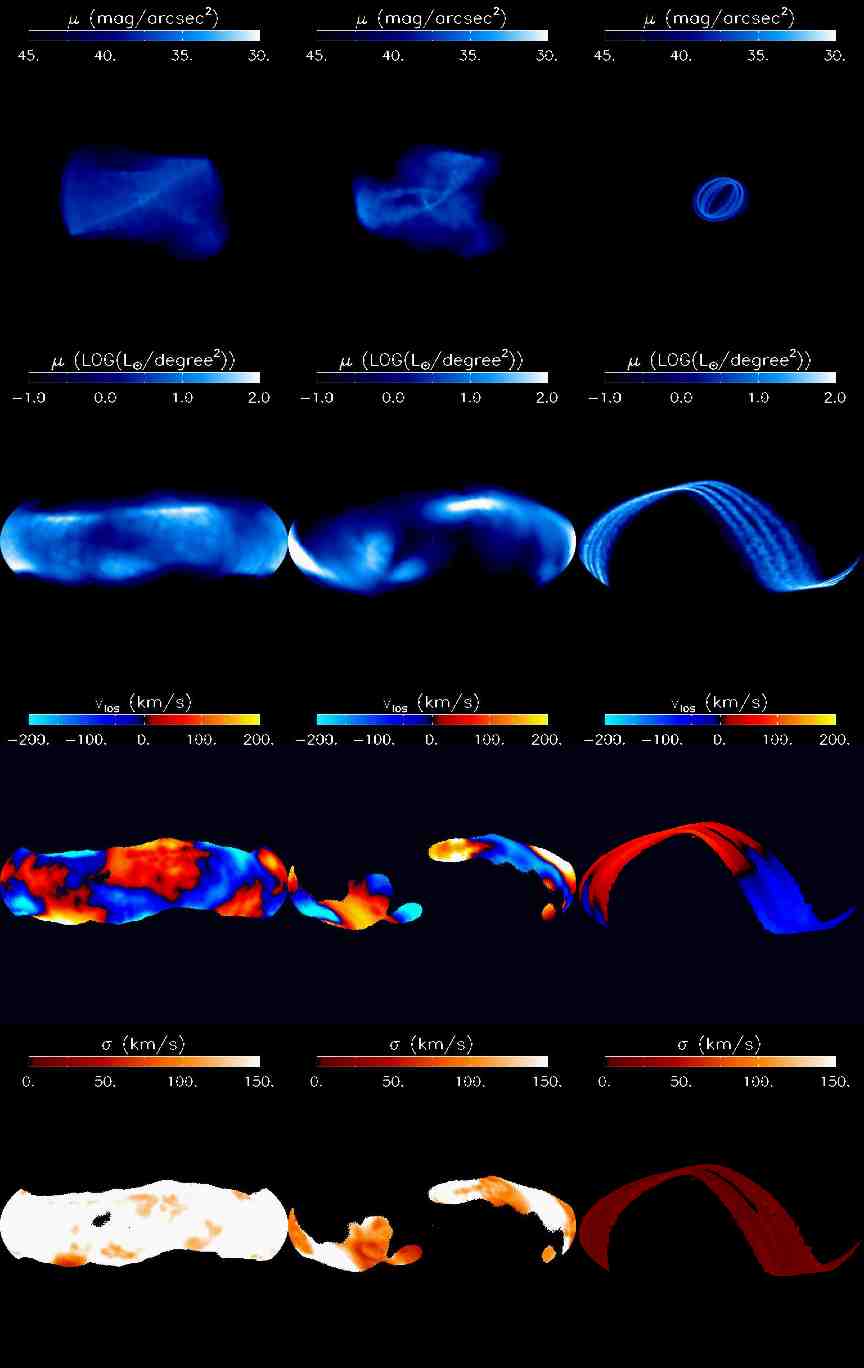}
\caption{\label{morph246.fig}
As Figure \ref{morph135.fig} for transitions --- mixed-clouds (left hand panels), cloudy-great-circles (middle panels) and mixed-great-circle (right hand panels) --- between the three morphological classes}
\end{figure*}

We visually  inspected plots of the surface density of all 1515 of our accretion events from an internal perspective (i.e. an Aitoff projection of the material as viewed from the parent galaxy center). Considering only the 1362 cases where no bound core of particles remained (determined from their mutual gravity and velocities relative to each other) we found that debris could be broadly characterized as belonging to a small set of morphological classes. Figure \ref{morph135.fig} illustrates the three main types in the classification scheme we adopted. (Note that the method for creating this and all other images in the paper is described in the appendix.)
\begin{description}
	\item{``Mixed'' morphologies} (example in left-hand panels) show a nearly uniform distribution of debris, equally spread above and below the plane of the parent galaxy, with the maximum exploration of the latitude set by the initial orbital inclination. The average line-of-sight velocity at a given point is typically low, with no evidence of global velocity gradients. The local velocity dispersion is high. The external view of this type of debris typically shows a donut shape, indicating full-mixing of unbound material throughout the entire phase-space region available to that orbit. Debris with this morphology would contribute to the smooth background of the stellar halo rather than adding distinct substructures. 
	\item{``Cloudy'' morphologies} (example in middle panels) show a few (typically 2 or 3) distinct density maxima, each a few tens of degrees across and only loosely (if at all) connected to each other by streams of intervening material. There are strong velocity gradients across these ``clouds'', and they can have low local velocity dispersions. The external view reveals debris strung along eccentric orbits with pericenters close to the parent galaxy center. Plumes lead out from the central galaxy to shells of material spreading out at the apocenters of the orbits.
	\item{``Great circle'' morphologies} (example in right-hand panels) show nearly uniform debris distributions aligned with a single great circle on the sky (viewing the debris from a solar perspective leads to some distortion of this great circle alignment, although the continuity of debris is maintained). There can be gradients in the line-of-sight velocity along the great circle, but more gentle than those seen in ``cloudy'' morphologies. The local velocity dispersion can be low, or high if the debris has become multiply-wrapped. The external view shows debris spread evenly along only mildly eccentric orbits.
\end{description}
Figure \ref{morph246.fig} shows examples of  events that were categorized as ``transitions''  between our three main morphological classes --- mixed-clouds, cloudy-great-circles and mixed-great-circles in the left, middle and right panels respectively.

\begin{figure}[t]
\epsscale{1.0}
\plotone{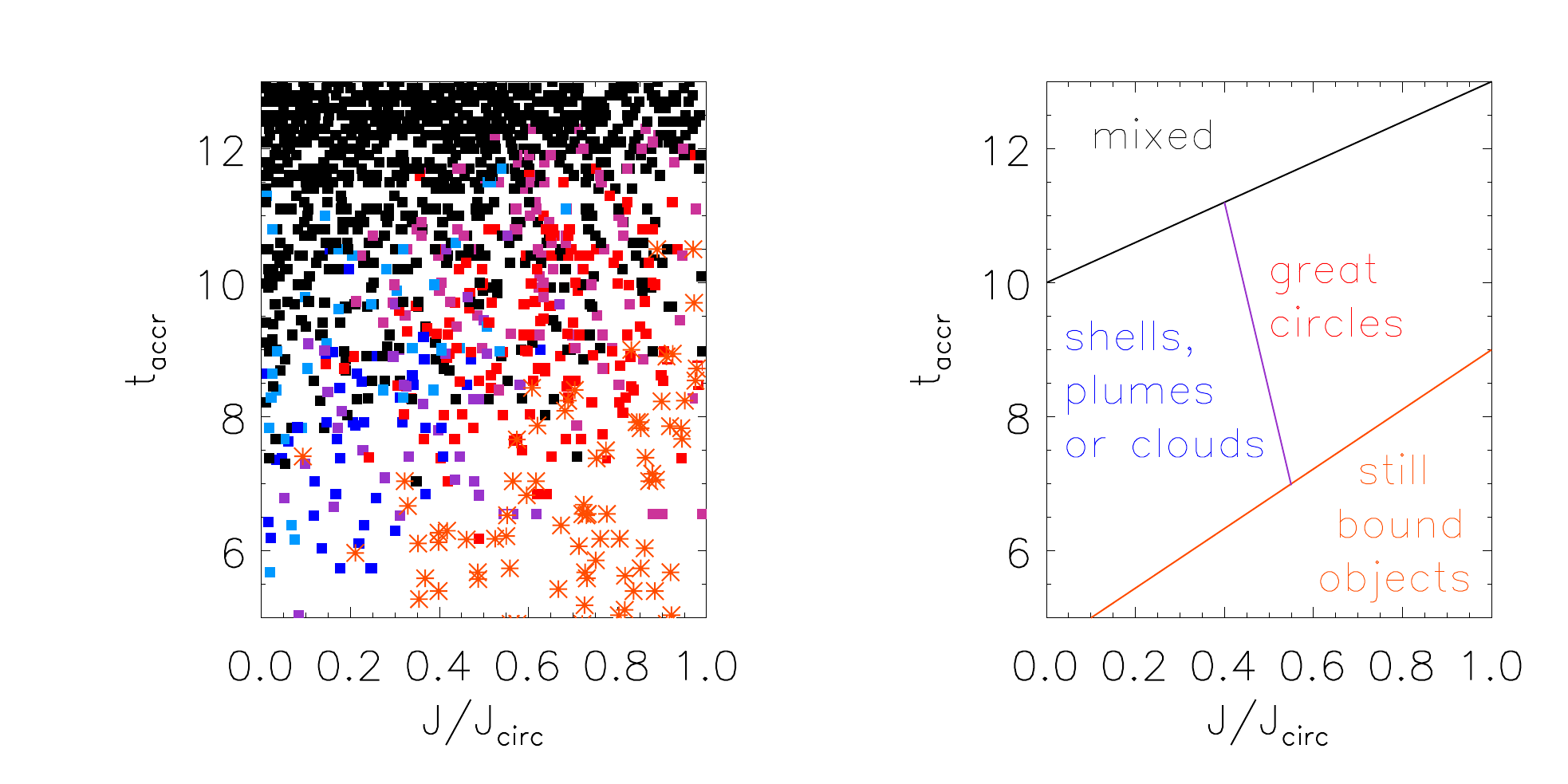}
\caption{\label{morphorb.fig}
Age {\it vs} $J/J_{\rm circ}$, color coded with morphology: blue and cyan indicate clouds and cloud/mixed morphologies; red and magenta indicate great-circles and great-circle/mixed morphologies; purple indicates cloud/great-circle morphologies; black indicates a mixed distribution; and orange stars are for still-bound satellites. The right-hand panel shows the dominant morphology contributing in each portion of this plane.}
\end{figure}

Figure \ref{morphorb.fig} summarizes the physical interpretation of our empirical scheme by plotting the accretion time {\it vs} orbital eccentricity (as characterized by $J/J_{\rm circ}$ -- the ratio of orbital angular momentum to the angular momentum of a circular orbit of the same energy)  for each of our accretion events, color coded with the assigned morphological classification. The right-hand panel illustrates our general conclusions: ``mixed'' morphologies arise from events accreted more than 10 Gyrs ago that have had time to fully mix along their orbits; ``clouds'' are from events accreted less than 8 Gyrs ago on eccentric (low $J/J_{\rm circ}$) orbits; ``great circles'' are from events accreted 6-10 Gyrs ago on near circular orbits ($J/J_{\rm circ}>$0.5); the transition types ``mixed-clouds'' and ``mixed-great-circles'' correspond to events that are partially, but not yet fully mixed along their orbits; and ``cloudy-great-circle'' morphologies correspond to recent events on moderately eccentric orbits.


\subsection{Surface brightness}
\label{mu.sec}


\begin{figure}
\epsscale{1.0}
\plotone{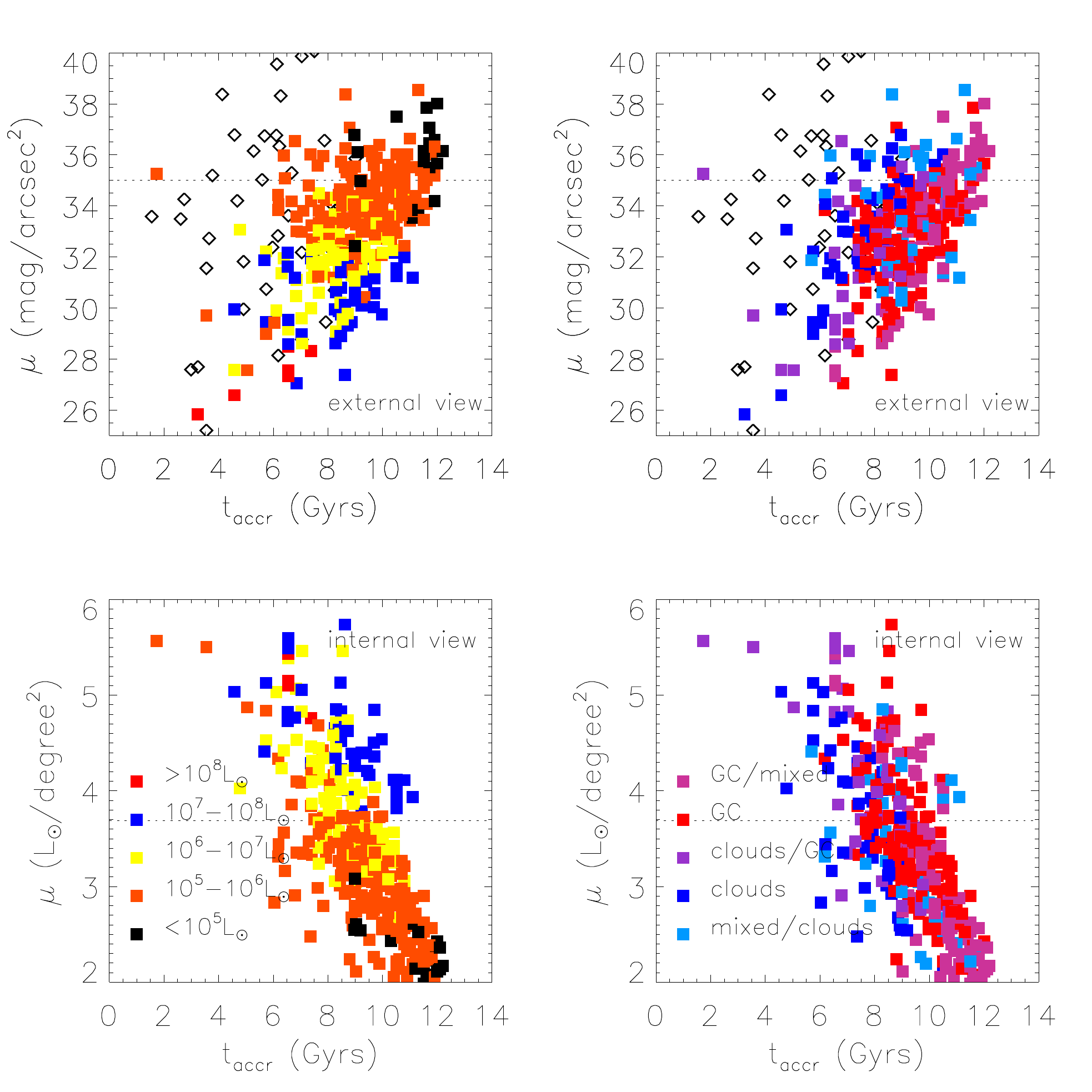}
\caption{
\label{mu.fig}
Maximum surface brightness of debris associated with accretion events in all eleven halos as a function of accretion time. Upper panels are from an external perspective and lower panels are from an internal perspective. Open diamonds are for debris from still-bound satellites. Black/orange/yellow/blue/red points in left-hand panels are for objects with luminosities $<10^5/10^5-10^6/10^6-10^7/10^7-10^8/>10^8 L_\odot$. Right-hand panels are color-coded with morphology using the same scheme as Figure \ref{morphorb.fig}.}
\end{figure}


Figure \ref{mu.fig} shows the maximum surface brightness of debris associated with events of a given accretion time from both external (upper panels) and internal (lower panels) viewpoints. 
This maximum was selected by looking along lines-of-sight to a random set of 10\% of the luminous particles in each simulation, and estimating the surface brightness from the total luminosity and sky coverage of their 30 nearest neighbors.  
(Using too few neighbors gave noisier results, and using too many systematically decreased the surface brightness estimates.)
In this and all subsequent plots, since we are interested in features that are likely to stand out distinctly from the parent galaxy's main components (i.e. be observable!), we only consider material at greater than 30kpc projected separation in the external view, or with distance moduli in the range than 15-21 ($\sim$ 10-160 kpc) in the internal view. For the same reason, we do not consider any events that were classified as purely ``mixed'' morphology (transition types are included).  
 
The left-hand panels of Figure \ref{mu.fig} are color-coded according to the satellite luminosity at the time of accretion. As might be expected in a hierarchical formation context there are few high-luminosity events at early times. Nevertheless, for a given accretion time there is a clear trend for the higher luminosity events to result in higher surface brightness-debris 
and in general the more recent events have higher surface brightness  \citep[as a consequence of the shorter phase-mixing time --- see][for more in-depth discussion of timescales for phase-mixing of debris.]{johnston98,helmi99a,johnston01} We conclude that the most obvious debris features observed around galaxies today should come from the most recent and most luminous accretion events.

The right-hand panels of Figure \ref{mu.fig} are the same as the left-hand panels, but color-coded according to the morphological type of the debris. As shown in \S \ref{morph.sec}, the transition mixed-clouds and mixed-great-circle types arise from earlier accretion events and hence have lower surface brightnesses. For events accreted at a given time, those with great-circle morphologies tend to have higher surface brightness. Nevertheless, the highest surface brightness events overall are predominantly cloud or cloudy-great-circle morphology. This is because our sample contains no events of great-circle morphology that were disrupted within the last 6 Gyears --- the progenitors of such debris would be recently accreted satellites on near-circular orbits and these are still bound in our simulations.

\subsection{Phase-space scales}
\label{ps.sec}


\begin{figure}[t]
\epsscale{0.5}
\plotone{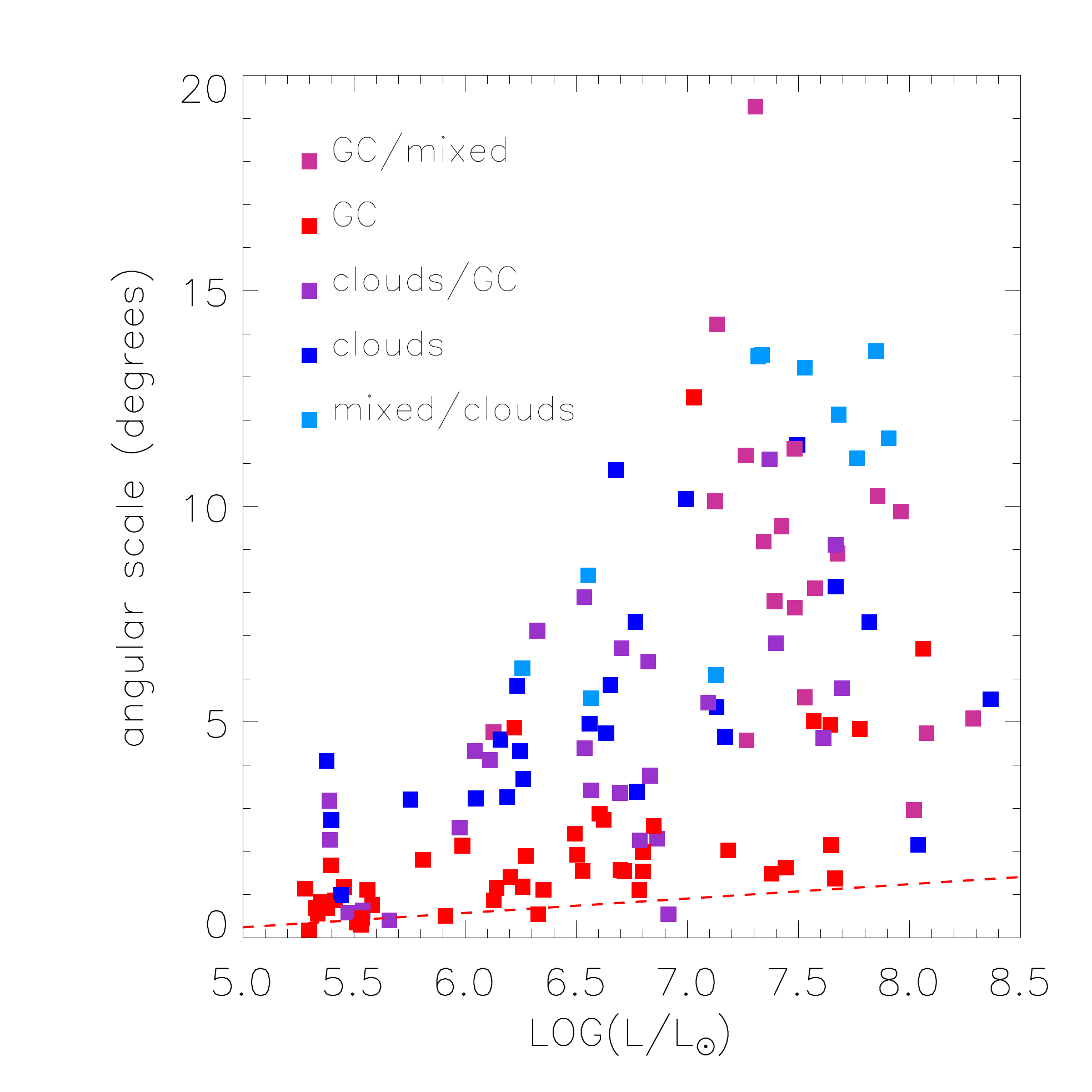}
\caption{
\label{psscale.fig}
Angular scales around the highest surface brightness point in the debris, color coded with morphology using the same scheme as in Figure \ref{morphorb.fig}. The red dotted line indicates the scaling given by equation (\ref{tidalscale.eqn}).
}\end{figure}


Previous studies of satellite disruption along mildly eccentric orbits show that the range in energy, angular momentum and orbital inclination in debris are a function of the progenitor satellite mass $m_{\rm sat}$ via the tidal scale 
\begin{equation}
\label{tidalscale.eqn}
	s=\left(m_{\rm sat}/M_{\rm Gal}\right)^{1/3},
\end{equation}
where $M_{\rm Gal}$ is the mass of the parent galaxy enclosed within the pericenter of the satellite's orbit \citep[e.g.][]{johnston98,warnick08}. 
The range in orbital energies corresponds to a range in orbital time periods, and this leads to the debris spreading along the orbit to form tidal streams.
The range in angular momentum corresponds to a range in the precession rate of the ``petals'' of the rosette orbits, and this leads to thickening of the streams in the orbital plane.
The range in orbital inclination corresponds to a range in the precession rates of the orbital plane, and leads to additional thickening of the streams perpendicular to the orbital plane.
The net effect of all this spreading is that the streams become less dense and (locally) colder over time \citep[as noted in \S 1, and see][]{helmi99a}. 
On the other hand, once the streams are multiply-wrapped (i.e. have spread more than $\pm\pi$ in azimuthal phase along the orbit) it is possible for an apparently local sample (i.e. selected spatially) to contain debris from several distinct orbital phases and have a large dispersion (e.g. see bottom panels of Figure \ref{morph135.fig}).

Figure \ref{psscale.fig} illustrates some of these ideas by plotting the angular scale of debris as viewed from the center of the parent galaxy for each of our accretion events, color-coded by debris morphology (and again omitting ``mixed'' morphologies). 
In cases of ``great-circle'' or ``great-circle/mixed'' morphology (red or pink points) the scale represents the width of the best-fit great circle containing 25\% of the debris. 
The width of debris with ``great-circle'' morphology roughly follows the expected scaling  (i.e. $\propto s$ --- see equation [\ref{tidalscale.eqn}]) indicated by the red line, even though variations in the mass-to-light ratio of the different objects (due to differing masses and accretion times) means that the translation between mass and luminosity is not strictly linear.
The pink points (``cloudy-great-circles'') fall systematically above this line because these events have had longer to spread beyond this width. If the precession of the orbital plane precession is small enough, their large width could also be due to multiple-wrapping in addition to the steady spreading that goes along with mixing. 

The blue points represent events with ``cloudy'' (dark blue) and ``cloudy-mixed'' (cyan) morphologies. In these cases the angular scale represents the radius containing 25\% of the light from the highest-surface-brightness cloud. This radius clearly follows a different (and generally larger) scaling than the great-circle width, indicative of a different mechanism for the origin of debris on eccentric rather than circular orbits (mass loss via sudden shocking rather than steady stripping). The trend for the older debris (as indicated by the more mixed ``transition'' type morphologies) to be more spread-out remains.


\begin{figure}[t]
\epsscale{1.0}
\plotone{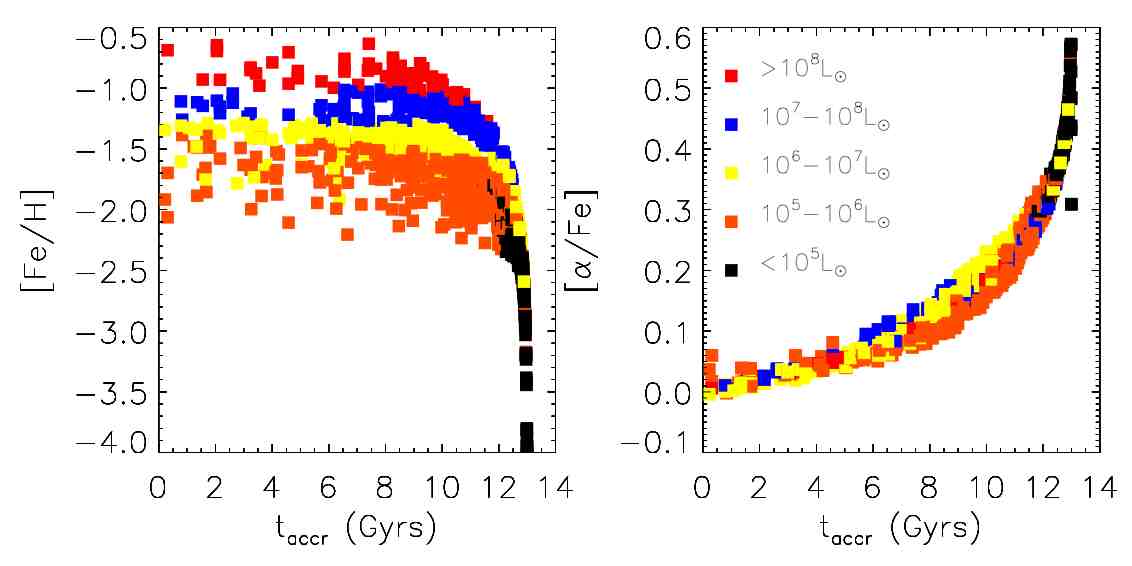}
\caption{
\label{fe.fig}
Mean [Fe/H] (left hand panels) and [$\alpha$/Fe] (right hand panels) of debris associated with accretion events in all eleven halos as a function of accretion time. Black/orange/yellow/blue/red points in left-hand panels are for objects with luminosities $<10^5/10^5-10^6/10^6-10^7/10^7-10^8/>10^8 L_\odot$. }
\end{figure}



\subsection{Stellar populations}
\label{fe.sec}

Figure \ref{fe.fig} plots the average [Fe/H] (left hand panel) and [$\alpha$/Fe] (right hand panel) as a function of the accretion time for each satellite, with the points color coded according to satellite luminosity. The main trends in the figure can be understood by recalling our assumptions that a satellite falling in today should sit on the observed mass-metallicity relation for dwarfs (hence lower luminosity dwarfs at any epoch have lower [Fe/H], as seen in the left-hand panel), and that all star formation ceases once a dwarf falls into the parent galaxy. The latter assumption means that a satellite's accretion time sets the length of its star formation history: those accreted long ago will have short histories, with their abundance patterns dominated by contributions from supernovae Type II (hence high [$\alpha$/Fe]); those accreted more recently will have long histories, with their abundance patterns containing by contributions from both supernovae Type II and supernovae type IA (hence low [$\alpha$/Fe], as seen in the right-hand panel). These ideas are discussed in more depth in \citet{robertson05}.

Note that the chemical properties of recently accreted material in our models (and substructure and surviving satellites in particular) are very well constrained since they depend on characteristics of model satellites that are fixed by requiring agreement with observations of Local Group dwarfs. 
However, the properties of earlier accretion events are likely to be less accurately reproduced since they depend on chemical evolution that is an extrapolation of current observational constraints.
In particular, while the general trends in abundance patterns as a function of accretion time and satellite mass seen in Figure \ref{fe.fig} can be thought of as fairly robust, the absolute values and scatter are indicative only and should be treated with some caution.

\subsection{Debris associated with surviving satellites}

\label{survivors.sec}

We deferred discussion of debris associated with still-bound systems since these represent a particular subset of events whose properties have been selected by dynamical shaping: the stars in the left-hand panels of Figure \ref{morphorb.fig} show that surviving satellites mostly come from recent events on mildly-eccentric orbits \citep[as seen in][]{font06a}. 
In addition, of the 153 survivors in our sample, only 33 have lost more than 1\% of their luminous matter and only 13 have lost more than 10\%, so debris from these objects is not typically a major contributor to stellar halos.

As might be expected from this summary, visual inspection of images to classify the morphology of debris from survivors revealed no debris for the majority of cases.
Of the remainder 18 had debris localized around the satellite, 22 had great-circle morphologies and 6 had cloudy-great-circle morphologies.
The open diamonds in the upper panels if Figure \ref{mu.fig} show the maximum surface brightness of debris from still-bound satellites is generally lower surface brightness than that from completely disrupted satellites that were accreted at the same time. Nevertheless, in our sample of 11 halos there are six still-bound satellites with associated debris brighter than 30th mag/arcsec$^2$. 

Overall, these results suggest that while finding a single satellite clearly in the process of disruption with debris spread around the parent galaxy should not be surprising in sufficiently sensitive surveys, finding many examples of satellites in this state around a single galaxy would be unusual.
Moreover, since surviving satellites tend to be on mildly-eccentric orbits we might expect debris associated with Galactic satellites to mostly be along great circles --- indeed the debris streams from the disruption of the Sagittarius dwarf galaxy have just such a morphology \citep[see e.g.][]{majewski03}.

\section{Results II: characteristics of substructure in stellar halos}

\label{resultsii.sec}

In this section, we build on the previous results to examine how the characteristics of substructure in stellar halos depend on the general properties of their constituent satellites --- i.e their merger histories.
Note that we only include contributions from unbound systems in our analyses --- although surviving satellites can contribute debris we do not expect a significant fraction of either the total content or number of substructures within a stellar halo to come from survivors (see \S \ref{survivors.sec}), so this simplification should not affect our results.

The results of \S \ref{morph.sec}-\ref{ps.sec} already emphasize the fundamental limitation of using coordinate-based studies to probe accretion histories back beyond the last few Gyears to the epoch when the most rapid growth is likely to have occurred for Milky-Way-type galaxies (i.e. $>$ 8 Gyears ago): older debris is fully mixed, morphologically indistinct and low surface brightness, hence hard to attribute to individual events and interpret in terms of accretion histories.
However, Figure \ref{fe.fig} clearly demonstrates that the early accretion epoch is precisely when we expect the most rapid evolution in abundance patterns to occur and these can provide a complementary probe of the histories at early times.


\begin{figure}[t]
\plotone{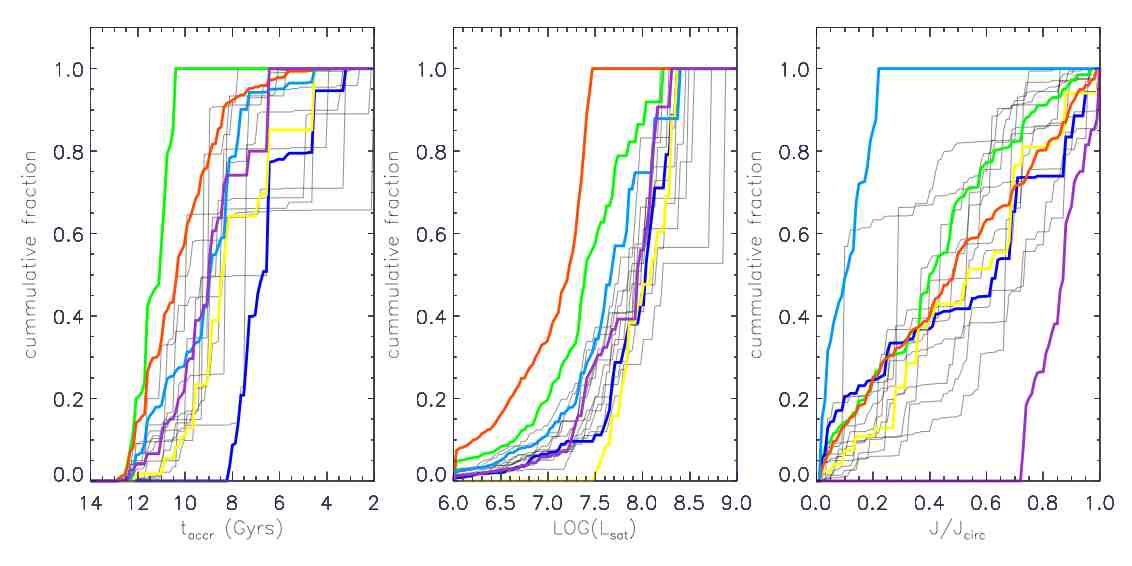}
\caption{\label{buildup.fig}
Fractional contribution of satellites to the total luminosity of the stellar halo as a function of accretion time, satellite luminosity and orbit type. Black lines are for our eleven standard halos. Halos built from ancient/recent events are shown in green/blue. Halos built from high/low luminosity events are shown in yellow/orange. Halos built from events on radial/circular orbits are shown in cyan/purple.}
\end{figure}


The black lines in Figure \ref{buildup.fig} illustrate the merger histories of our eleven ``standard'' halos by plotting the fractional contribution of events to the total luminosity of the halo as a function of accretion time, satellite luminosity and satellite orbit. 
The bulk of these halos are built from accretion events occurring more than 8 Gyears ago, with luminosities greater than several times $10^7 L_\odot$ and on a mixture of radial and circular orbits.

In order to explore how sensitive the appearance of substructure is to different merger histories, six ``artificial'' model halos (each containing roughly $10^9L_\odot$ in stars) were constructed by simply summing over accretion events from our entire  library chosen to have the required properties:
\begin{itemize}
\item
our ``ancient'' and ``recent'' halos (green and dark blue lines in Figure \ref{buildup.fig}) were built from events entirely accreted longer than 11 Gyears ago or more recently than 8 Gyears ago;
\item
our ``high luminosity'' and ``low luminosity'' halos (yellow and orange lines in Figure \ref{buildup.fig}) were built from events either more or less luminous than several times $10^7 L_\odot$;
 \item
our ``radial'' and ``circular'' halos were built from events predominantly on radial or circular orbits (with $J/J_{\rm circ}<0.2$ or .$J/J_{\rm circ}>0.75$ --- cyan and purple lines in Figure \ref{buildup.fig}). 
\end{itemize}
Of course, while these histories clearly lie outside the region in mass, orbit and accretion-time expected for the majority of galaxies in a $\Lambda$CDM cosmology (as indicated by the black lines), it is not impossible that some galaxies in the Universe were formed in just this way. 

\subsection{Number of features} 

\label{frequency.sec}

\begin{figure}[t]
\plotone{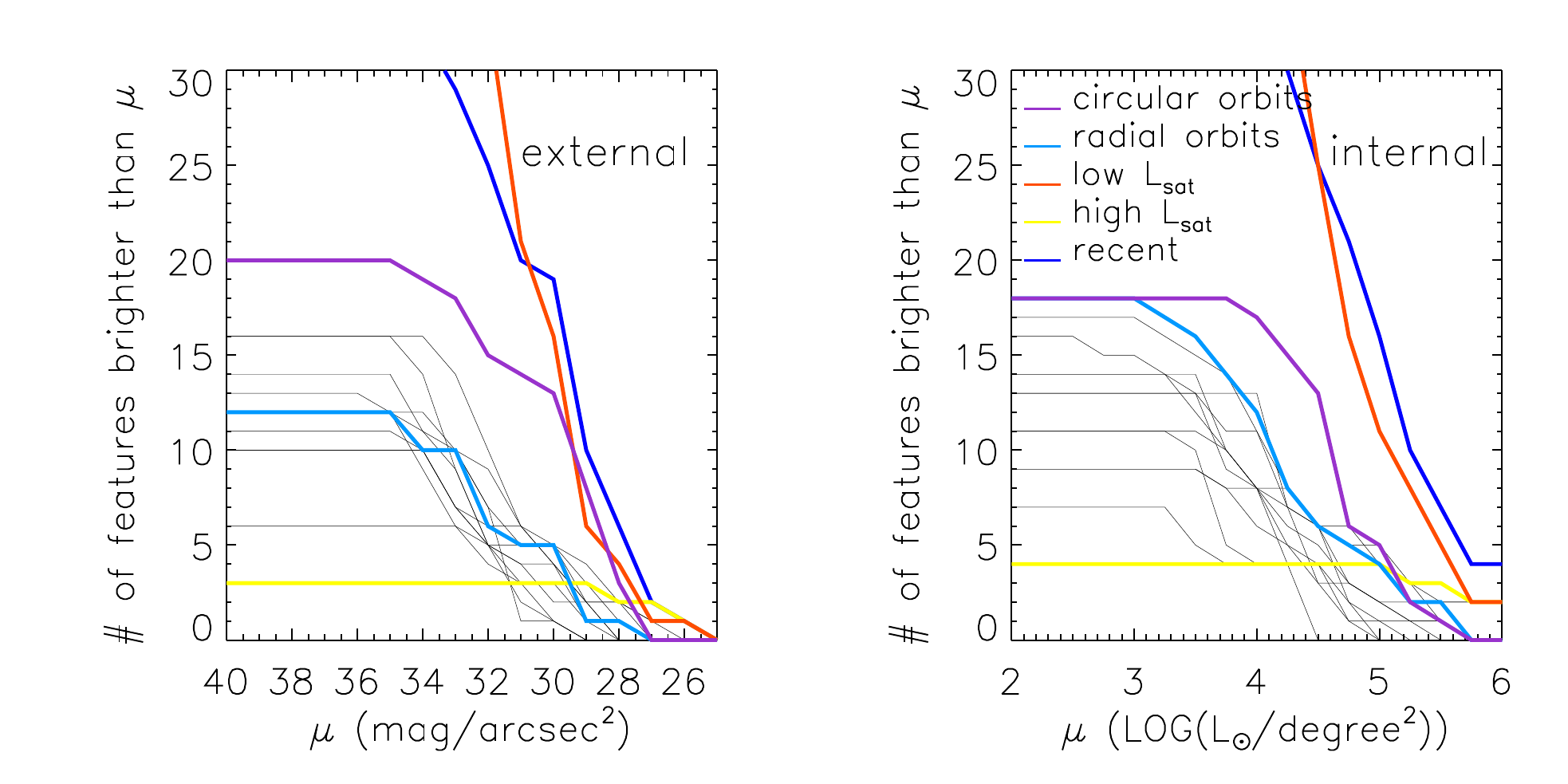}
\caption{\label{frequency.fig}
Left hand panel: Number of accretion events with debris above the background (see text) and brighter than $\mu$ at projected distances greater than 30kpc from the host.  Right hand panel: the same thing from an internal perspective for debris with distance moduli in the range 15-21.
Black lines are for our eleven standard halos. Halos built from recent events are shown in blue. (Note that all events in our ``ancient'' halos resulted in debris with diffuse morphology that was not counted in this figure.) Halos built from high/low luminosity events are shown in yellow/orange. Halos built from events on radial/circular orbits are shown in cyan/purple.}
\end{figure}


We estimated the number of substructures that could be seen around our model galaxies by counting all events that were not of ``mixed'' morphology and contributed more than 50\% of the total stellar halo light at their brightest point (i.e. to be considered above the background).  
Figure \ref{frequency.fig} shows the cumulative number of satellites contributing features brighter than $\mu_{\rm max}$ in all eleven ``standard''  halo models (black lines) as well as our six artificial halo models from both external (left hand panel) and internal (right hand panel) viewpoints.  

The black lines show that distinct debris features brighter than 26th mag/arcsec$^2$ should be unusual, while surveys reaching to 30th/35th mag/arcsec$^2$ can be expected to see of order a few to a dozen features around Milky-Way type galaxies. The simplified nature of our modeling is likely to prolong the clarity of substructure, which suggests that the lines in Figure \ref{frequency.fig} should be considered upper limits on the number of satellites likely to be contributing. However, our analysis considers only the brightest point of each event's debris and in reality debris from a satellite might contribute several distinct observable features. Hence these numbers are best thought of as order-of-magnitude estimates for what we might see.
\citep[Note that these numbers are rather greater than the lower limits estimated by][in part because these authors only considered events that were accreted after the parent galaxy had grown to 90\% of its current size.]{johnston01}

The colored lines give us some idea of what this frequency plot is telling us.  The blue lines show that there are many more features around our ``recent'' halo than seen in our ``standard'' halos since debris from recent events tends to have distinctive morphology and higher surface brightness. Our ``low luminosity'' halo (orange lines) has a comparable number of features to our ``recent'' halo simply because of the sheer number of low luminosity accretion events required to put it together - it has a similar number of accretion events in the last 7 Gyrs as our ``recent'' halo even though the bulk of its mass was accreted much earlier. The opposite is true for our ``high luminosity'' halo (yellow lines). The number of features around our ``radial'' and ``circular'' halos falls within the distribution of our eleven ``standard'' halos because the accretion histories for these halos are close to the standard ones. Finally, there are no green lines (corresponding to our  ``ancient'' artificial halo) because the satellites randomly selected to build it from had debris all of ``mixed'' morphology --- the lack of recent accretion means that all events have had time to mix along their orbits and leave no distinctive features. Overall, Figure \ref{frequency.fig} suggests that  the number of features around a galaxy can be broadly related to the number of recent accretion events.

\subsection{Fraction of material in substructure}
\label{amount.sec}


\begin{figure}[t]
\plotone{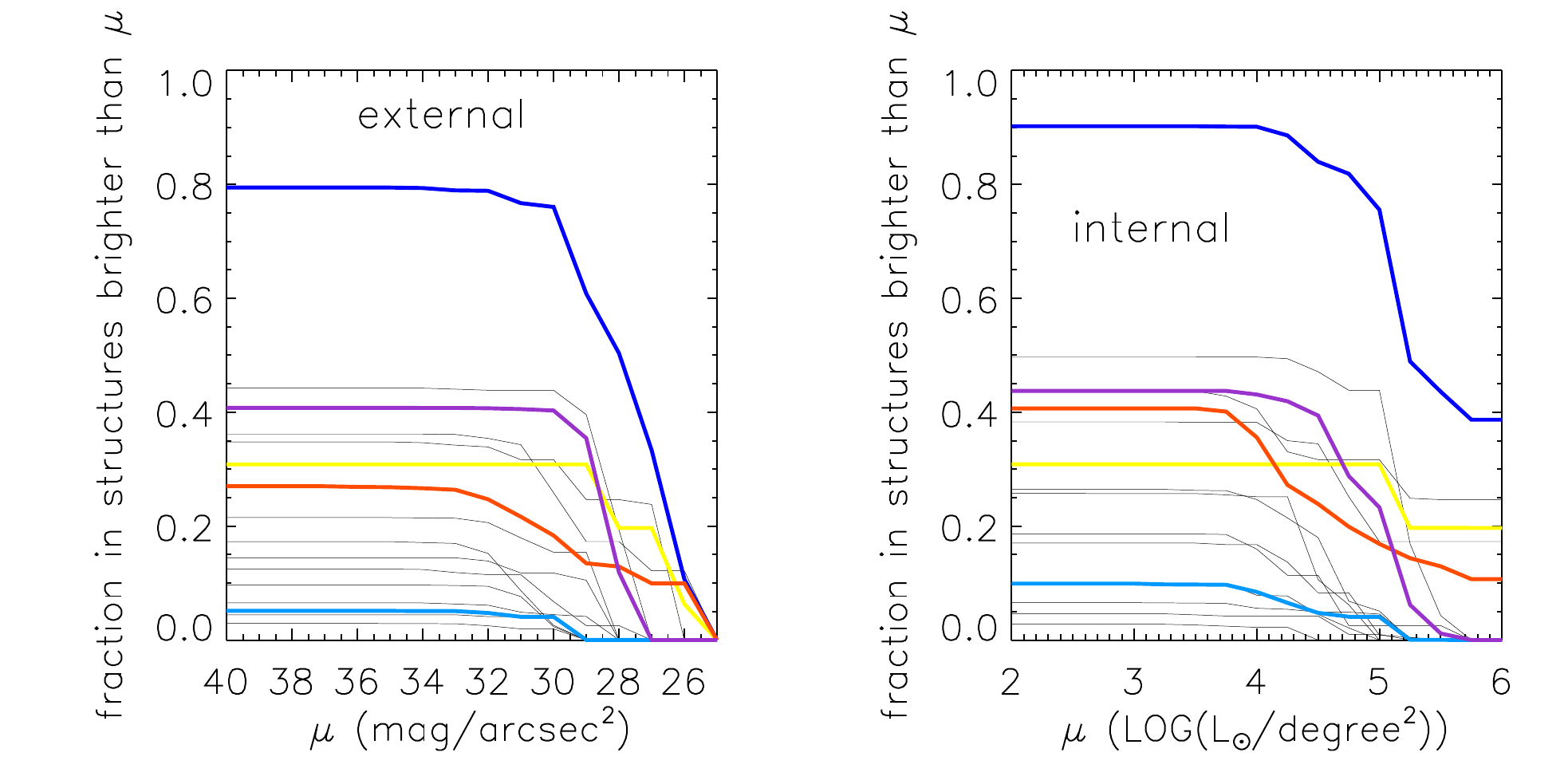}
\caption{\label{amount.fig}
Fraction of halo luminosity contributed by objects with associated debris brighter than $\mu$ and above the background (as defined in the text). Black lines are for our eleven standard halos. Halos built from recent events are shown in blue. (Note that all events in our ``ancient'' halos resulted in debris with diffuse morphology that was not counted in this figure.) Halos built from high/low luminosity events are shown in yellow/orange. Halos built from events on radial/circular orbits are shown in cyan/purple.}
\end{figure}


The amount of material in substructure is assessed in Figure \ref{amount.fig} by plotting the cumulative fraction of the stellar halo contributed by the satellites with debris of non-uniform morphology (i.e. counted in Figure \ref{frequency.fig}) brighter than surface-brigthness $\mu$. 
The black lines show that we should in general expect $\sim$10\% of the stars in a stellar halo to be in the form of distinct features, with a large range (1-50\%) around this typical value. 
Although there are more features by number
fainter than $\mu=$30 mag/arcsec$^2$ (see Figure \ref{frequency.fig}) , a larger fraction of the material in substructures in the stellar halo are associated with features brighter than this. 

The colored lines in Figure \ref{amount.fig} suggest that the primary factor determining the amount of material in substructure is the epoch of accretion. The clear outliers in Figure \ref{amount.fig} are the blue and green lines which correspond to  ``recent'' and ``ancient'' halos respectively, while those with more usual accretion epochs fall within the range of our ``standard'' models. In particular, note that even though Figure \ref{frequency.fig} shows that there are similar numbers of features in the ``recent'' and ``low-luminosity'' halos the former contains 80-90\% of its material in substructure, while the latter has a more typical value of $\sim$25\%.

\subsection{Morphology of features}


\begin{figure}[t]
\plotone{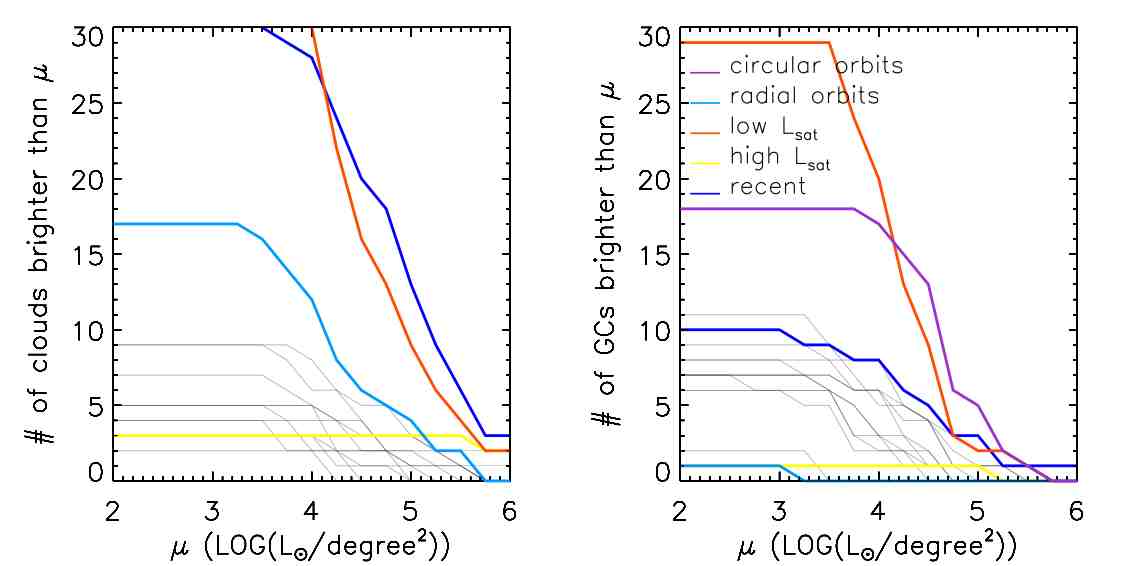}
\caption{\label{freqmorph.fig}
Repeats right-hand panel of Figure \ref{frequency.fig} for features of mixed/cloud, cloud or cloud/great-circle morphologies (left hand panel) and great-circle or great-circle/mixed morphologies (right hand panel). Black lines are for our eleven standard halos. Halos built from recent events are shown in blue. (Note that all events in our ``ancient'' halos resulted in debris with diffuse morphology that was not counted in this figure.) Halos built from high/low luminosity events are shown in yellow/orange. Halos built from events on radial/circular orbits are shown in cyan/purple. No purple lines appear in the left-hand panel because no debris with cloud morphology was found in the halo built from circular orbits.}
\end{figure}


Figure \ref{freqmorph.fig} repeats the right hand panel of Figure \ref{frequency.fig} (the frequency of features in general) for debris classified as having cloud (left hand panel -- including mixed/cloud, cloud and cloud/great-circle types) and great-circle morphologies  (right hand panel  -- including great-circle and great-circle/mixed  types). 
Note that  there are roughly equal numbers of cloudy and great-circle morphologies for features in our standard halos (comparing the black lines in the left- and right-hand panels) and low- or high-luminosity halos (orange and yellow lines). Our stellar halos built from radially-biased/circularly-biased orbits (cyan and purple lines respectively) show distinctive patterns with a larger fraction of debris in cloud/great-circle morphology . Hence the distribution of distinct debris morphologies in halos should tell us something about the distribution of orbital properties of the recently accreted satellites.

Our ``recent'' halo also appears to contain more clouds rather than great circle morphologies. This is a reflection of the fact that surviving satellites tend to be relatively recent accretion events in near circular orbits --- debris in the recent halo will tend to come from objects that were completely destroyed and be biased towards radial orbit events. 

\subsection{Radial distribution of substructure}


\begin{figure}[t]
\plotone{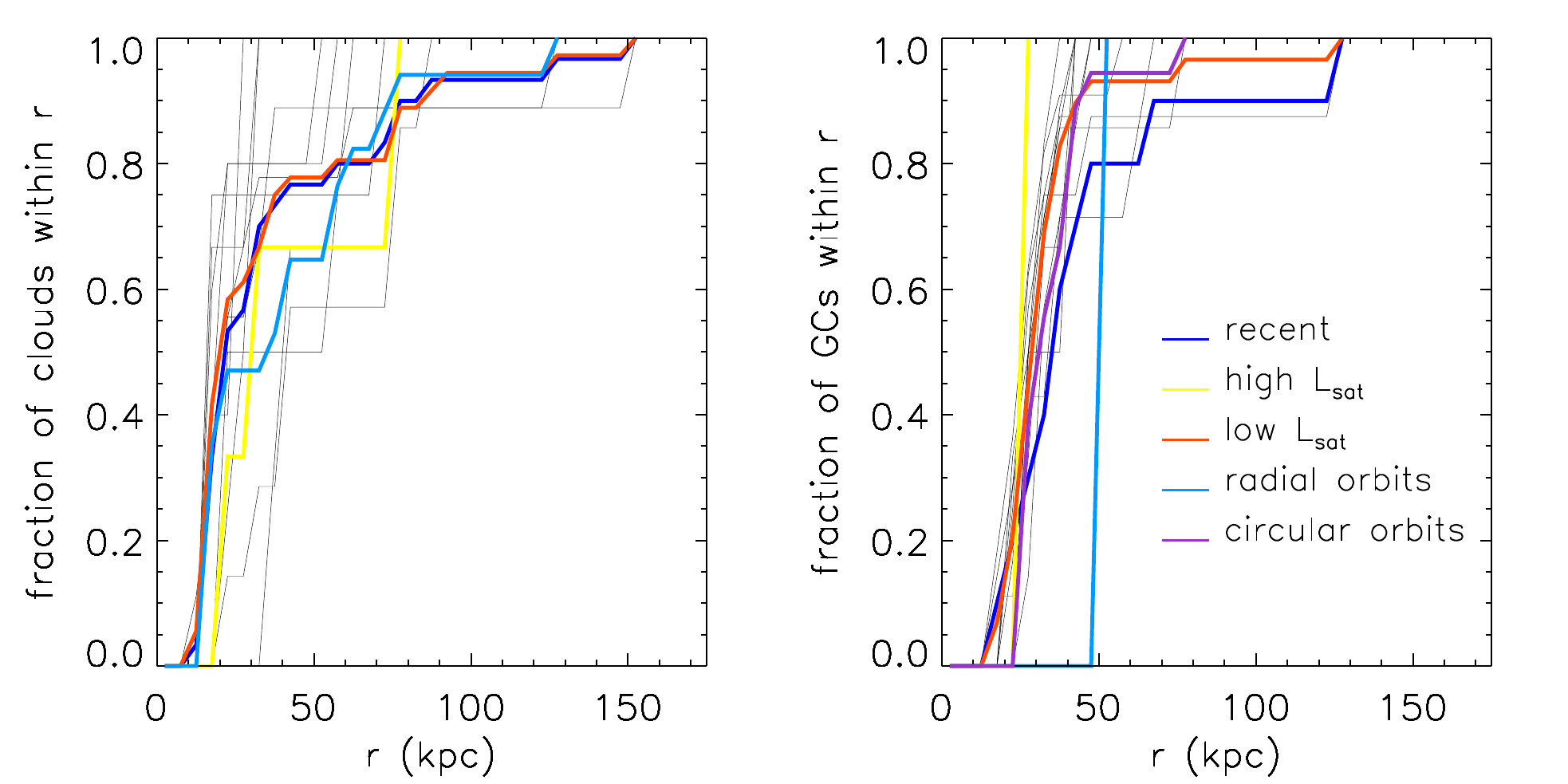}
\caption{\label{radius.fig}
Number fraction of maxima of cloud, mixed-cloud and cloud/great-circle morphologies (left-hand panels) or great-circle and great-circle/mixed morphologies (right hand panel) within distance $r$ of Sun.}
\end{figure}


Figure \ref{radius.fig} plots the number fraction of features with maxima within distance $r$ of the Sun, separately for features with ``cloud'' (left hand panel) and ``great circle'' (right hand panel) morphologies. Neither panel contains any features within 10kpc of the Sun, most of the great circles lie in the range 30-50kpc and the clouds occupy a larger range in distance (20-100kpc). Hence we expect stellar halos to be increasingly dominated by substructure as surveys explore larger and larger distances from the host galaxy, and at the largest distances we anticipate that most of this substructure would be in the form of clouds.

\subsection{Angular scales of features}


\begin{figure}[t]
\epsscale{0.5}
\plotone{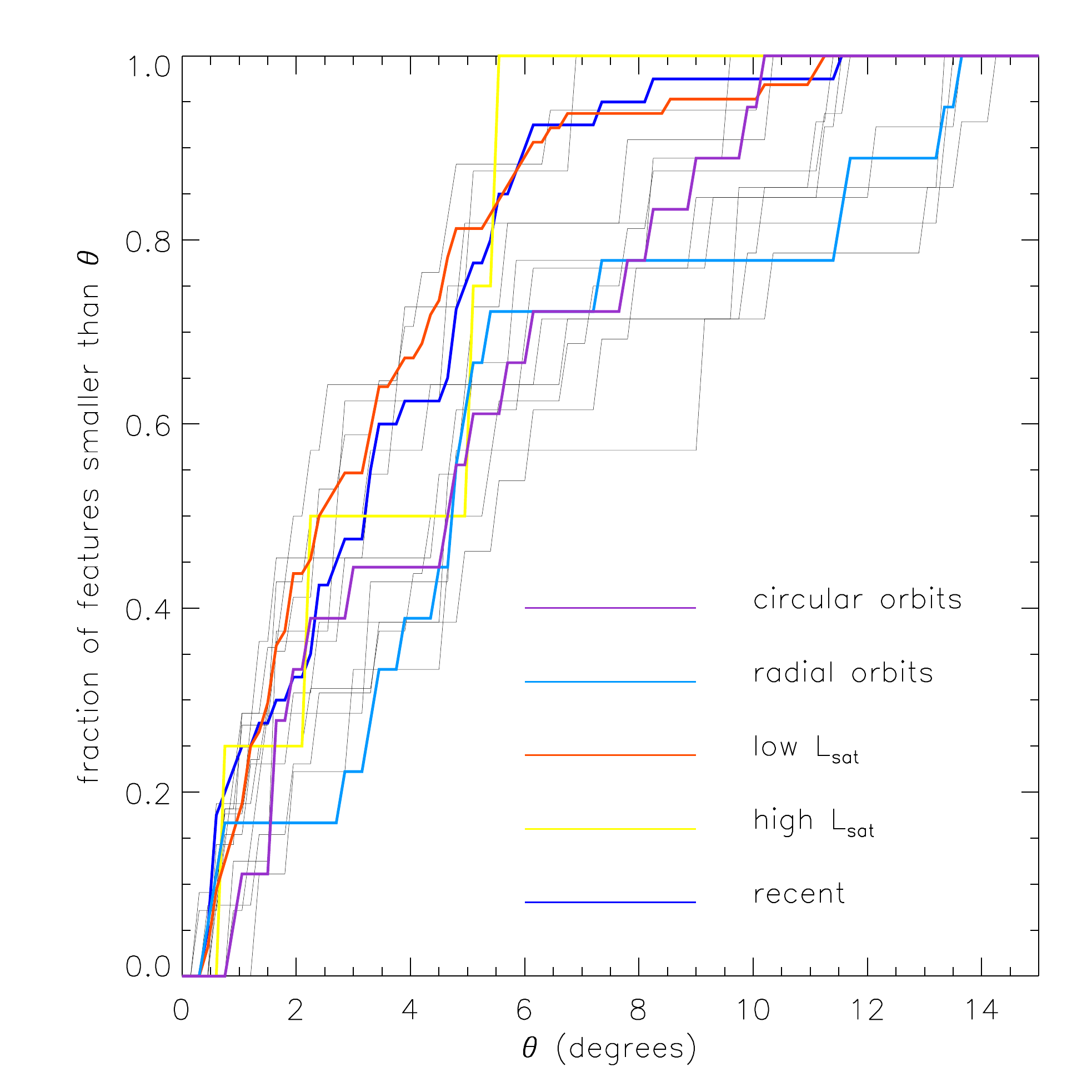}
\caption{\label{freqangle.fig}
Frequency of angular scales in debris, as seen from an internal perspective. Black lines are for our eleven standard halos. Halos built from recent events are shown in blue. (Note that all events in our ``ancient'' halos resulted in debris with diffuse morphology that was not counted in this figure.) Halos built from high/low luminosity events are shown in yellow/orange. Halos built from events on radial/circular orbits are shown in cyan/purple.
}
\end{figure}


Figure \ref {freqangle.fig} illustrates the distribution of angular scales for debris of non-uniform morphology (i.e. plotted in Figure \ref{frequency.fig}). In general, all our halos show a range of scales, from a few to ten degrees across. As might be anticipated from our understanding that debris with great-circle morphology tends to have smaller angular scales than debris with cloud morphology (see Figure \ref{psscale.fig})
the artificial halos built from circular orbits (purple line) have smaller angular scales than those built from radial orbits (cyan line). The halo built from low luminosity events (orange line) also has smaller scales than the typical standard halo. Hence the typical scales of substructure reflects both the luminosity function and orbital distribution of infalling objects.

\subsection{Stellar populations}

\label{abundances.sec}

\begin{figure}[t]
\epsscale{0.5}
\plotone{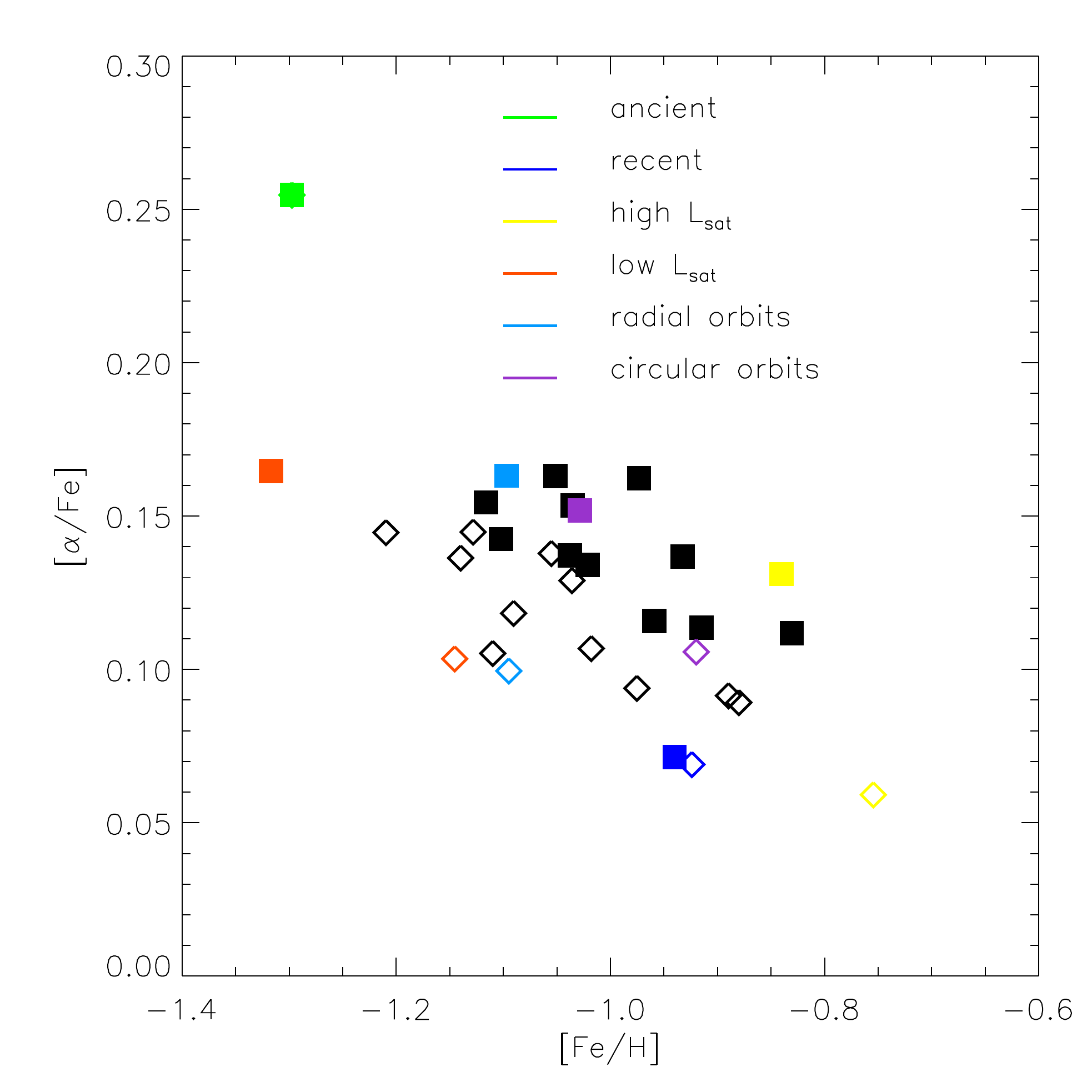}
\caption{\label{abundances.fig}
Luminosity-weighted average [$\alpha$/Fe]  against [Fe/H] for entire content  of each halo (solid points). The open points indicate luminosity-weighted averages for the events that contribute clear substructure (i.e surface brightness maxima above the background, as seen from an internal perspective).}
\end{figure}


Figure \ref{abundances.fig} plots the average [$\alpha$/Fe]  against [Fe/H] for the entirety of (solid points) and substructure within (open points) our stellar halos. In nearly every case, the substructure is chemically distinct from the halo as a whole. As noted in \citet{font06a,font06b} this distinction can be interpreted as corresponding to the distinction in accretion epoch of the stellar halo in general ($>$8 Gyrs ago) {\it vs} substructure ($<$ 8 Gyrs ago). The one exception to this trend is the ``recent'' stellar halo which was all accreted $<$ 8 Gyrs ago. 

The colored points show how this intuition can be be applied to interpreting abundance patterns in stellar halos in general. The green/blue points (ancient/recent halos) bracket the more general distribution of points in [$\alpha$/Fe], while the orange/yellow points (low/high luminosity halos) bracket the distribution in [Fe/H]. Hence, in our model halos the [$\alpha$/Fe] distribution  is telling us about the epoch of halo accretion, while the [Fe/H] distribution is sensitive to the luminosity function of accreted satellites. Conceptually, we can imagine splitting the [Fe/H]-[$\alpha$/Fe] plane into four quadrants corresponding to accretion histories dominated by ancient, high-luminosity events (high [Fe/H] and [$\alpha$/Fe]), ancient, low-luminosity events (low [Fe/H] and [$\alpha$/Fe]), recent, high-luminosity events (high [Fe/H] and low [$\alpha$/Fe]) and recent, low-luminosity events (low [Fe/H] and [$\alpha$/Fe]).

As noted in \S \ref{fe.sec}, the absolute values in the above results will depend on the assumptions in our models for how chemical enrichment within each dwarf proceeds. Nevertheless, the trends apparent in Figure \ref{abundances.fig} demonstrate the potential power of using abundances as a probe of early halo histories --- precisely the epoch that cannot be probed by debris morphology alone because it has had time to become fully mixed.

\section{Summary: interpreting halo properties in terms of accretion histories}
\begin{figure}
\epsscale{1.0}
\plotone{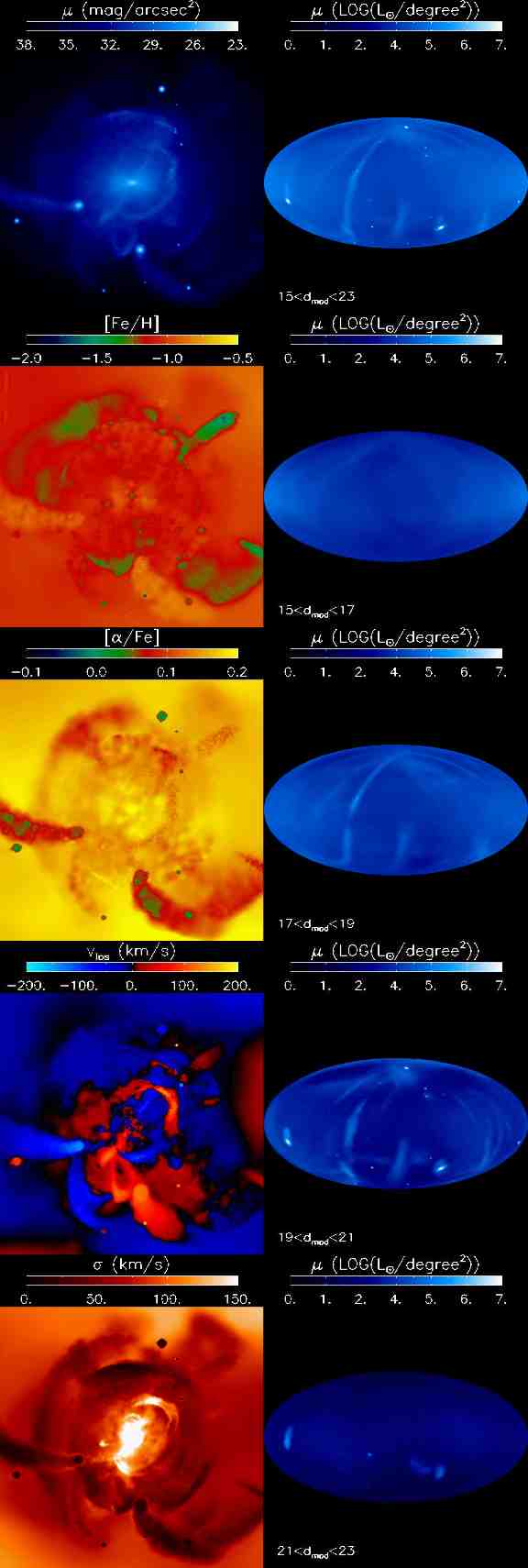}
\caption{\label{halo.fig}
Left hand column shows the surface brightness, average [Fe/H],  [$\alpha/$Fe], line-of-sight velocity and dispersion (panels running top to bottom)  for our standard halo number 8, as viewed from an external perspective. 
Each panel is 300kpc on a side.
Right  hand column shows all sky projections (as viewed from a point 8kpc form the center of the galaxy) of the surface density of stars within different distance moduli ranges. The top right-hand panel shows the full range of distance moduli considered (15-23, or 10-398 kpc)}
\end{figure}

\begin{figure*}
\epsscale{0.7}
\plotone{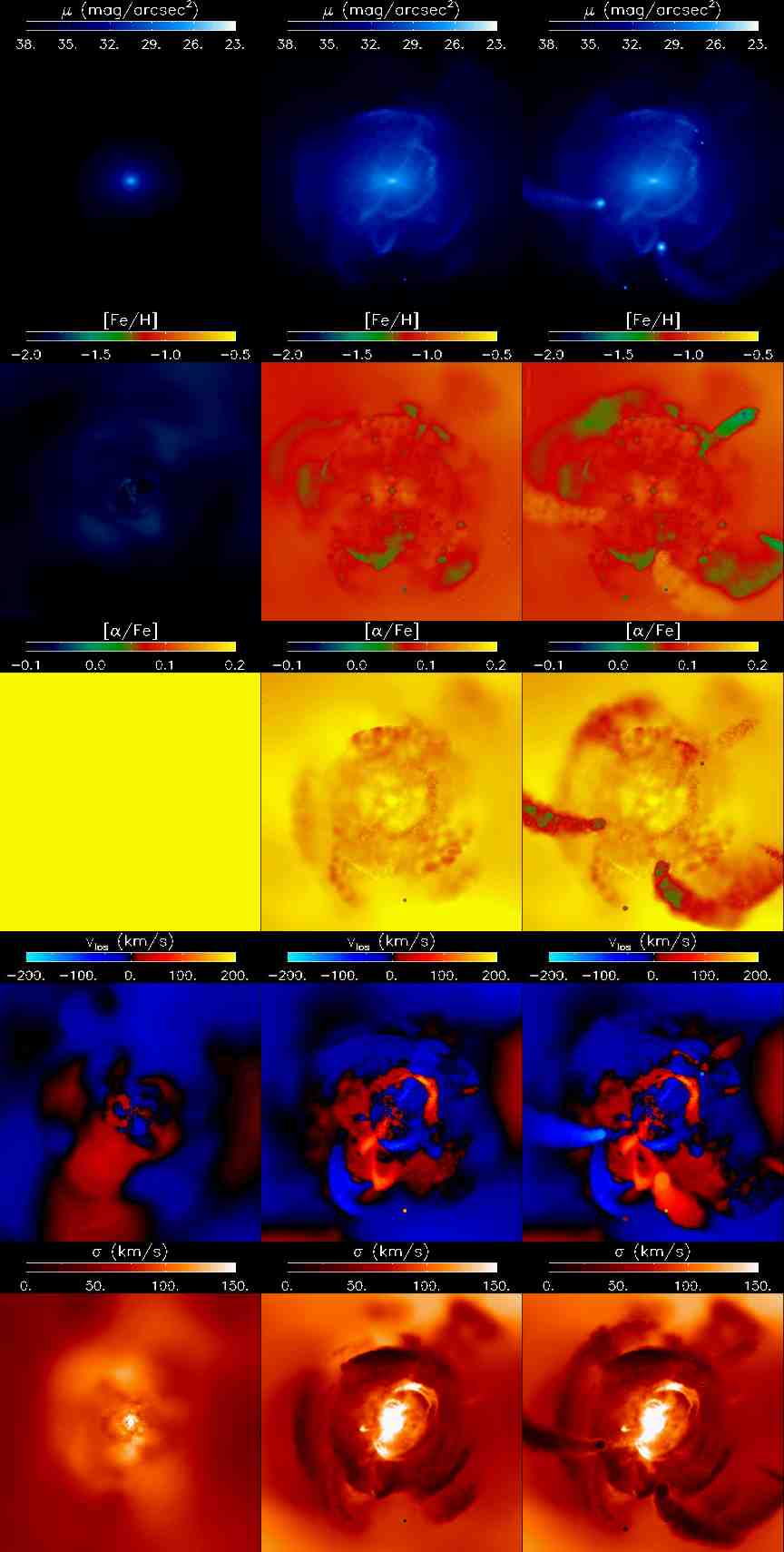}
\caption{\label{evol.fig}
Contribution to standard halo 8 in surface brightness, average [Fe/H],  [$\alpha/$Fe], line-of-sight velocity and dispersion (panels running top to bottom)  from all satellites accreted before 12/8/4 Gyrs ago (in columns from left to right). Each panel is 300kpc on a side.}
\end{figure*}

\label{summary.sec}

Having examined how the characteristics of stellar halos are related to the properties of their constituent satellites in \S \ref{resultsi.sec} and \S \ref{resultsii.sec} we now go on to synthesize and illustrate these results using phase- and abundance- space maps of all of our halos.

Figure \ref{halo.fig} summarizes our expectations for the general properties of stellar halos built within  a $\Lambda$CDM universe through images of one of our ``standard'' halos. The left hand column shows external views of surface brightness, average [Fe/H], [$\alpha$/Fe] and velocity, and velocity dispersion along the line-of-sight.
The right-hand column shows surface densities of stars as viewed form a point 8 kpc from the center of the galaxy in a set of shells of increasing radius. 
The figure illustrates that, for typical accretion histories of Milky-Way type galaxies, a stellar halo will show: of order 10\% of its content in substructure, with most of that substructure beyond $\sim$20kpc from its center; a handful of features brighter than 30th mag arcsec$^{-2}$ and dozens brighter than 35th;  roughly equal numbers of these features with cloud and great circle morphology; clouds (or plumes and shells) occurring over a larger range of  galactocentric distances (10-100kpc) than great circles (or rosettes --- 30-50kpc); $\sim$ 1 of its satellites with tidal tails brighter than 30th mag/arcsec$^2$; and abundance patterns that are distinct for stars in the smooth component ($\alpha$-rich), highest surface-brightness substructures (often metal rich and always $\alpha$-poor) and surviving satellite population (typically metal poor and $\alpha$-poor) of the halo.  

As an example of how these typical properties arise, Figure \ref{evol.fig} shows the contribution to the images in the left-hand column of Figure \ref{halo.fig} from satellites of different accretion times. Panels in columns 1/2/3 show contributions from all satellites accreted more than 12/8/4 Gyears ago. These columns illustrate our general intuition that the inner halo is built early on (more than 8 Gyears ago), and that the debris from these early events is now fully mixed (see Figure \ref{morphorb.fig}). The brightest substructures dominate the outer halo and typically come from more recent events (see Figure \ref{mu.fig}). Surviving satellites typically fall in only a few Gyears ago \citep[see][]{bullock05}. It is precisely because of their late infall that both satellites and dominant substructures are chemically distinct from the bulk of the stellar halo \citep{robertson05,font06a,font06b}. Satellites and substructures are often chemically distinct from each other because the former are dominated (by number) by low-luminosity (and hence low-metallicity) objects, while the brightest (i.e. first detected) examples of the latter are often from high-luminosity (and hence high-metallicity) objects \citep{font08,gilbert08}.

\subsection{General trends in halo properties with accretion histories}
\begin{table*}[t]
\begin{center}
\begin{tabular}{|c|c|c|}
\hline
Observable property				&	Interpretation		& Implication  \\ 
\hline
\hline
fraction in 						&	recent			& high fraction $\Rightarrow$ many recent events  \\   
substructure					&	accretions			& low fraction $\Rightarrow$ few recent events  \\
\hline
scales in						&	luminosity	function	& large $\Rightarrow$ high luminosity events \\	
substructure					&	(and orbit type) 	& small $\Rightarrow$ low luminosity events \\	
							&	of recent events &\\
\hline
number of 					&	number of			& large $\Rightarrow$ many events	\\
features						&	recent events		& small $\Rightarrow$ few events	\\
\hline
morphology					&	orbit				& clouds/plumes/shells $\Rightarrow$ radial orbits \\
of substructure					&	distribution		& great circles $\Rightarrow$ circular orbits \\
\hline
[Fe/H]						& luminosity 			& metal-rich $\Rightarrow$  high luminosity events \\	
							& function				& metal-poor $\Rightarrow$ low luminosity events \\
\hline
[$\alpha$/Fe]					&	accretion 			& $\alpha$-rich $\Rightarrow$ early accretion epoch \\
							&	epoch			& $\alpha$-poor $\Rightarrow$ late accretion epoch \\
\hline
\end{tabular}
\caption{Summary of general trends for stellar halo interpretation}
\label{trends.tab}
\end{center}
\end{table*}

In reality, there is a large range in halo properties around the general description given above, which can be interpreted as being due to the corresponding range in accretion histories.
The results of \S\ref{resultsi.sec} and \S\ref{resultsii.sec} clearly demonstrate that the frequency of substructure and the fraction of a stellar halo in substructure around a galaxy is sensitive to both the epoch when it accreted most of its mass and the luminosity scales of the objects it accreted, while the morphology of the substructure reflects the orbit distribution of the progenitor satellites.
Unfortunately, substructure in coordinate space only remains apparent in surface brightness and morphology for sufficiently long to probe the more recent accretion history of a galaxy in detail, corresponding to approximately the last 10\% of its mass growth in a $\Lambda$CDM universe. 
However, we expect the abundance patterns in stellar populations to evolve most rapidly at early times, and hence offer a complementary probe of the early accretion history of a galaxy. 
Table \ref{trends.tab} summarizes our understanding of phase- and abundance-space signatures of accretion histories developed so far.

\begin{figure*}
\epsscale{1.0}
\plottwo{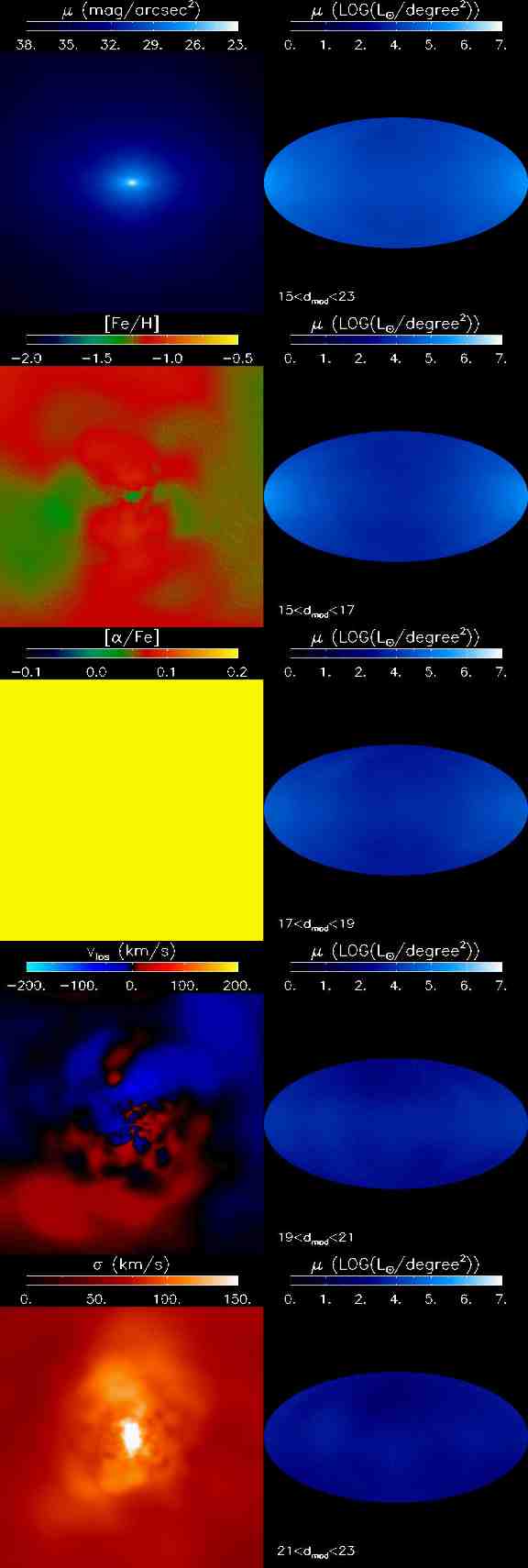}{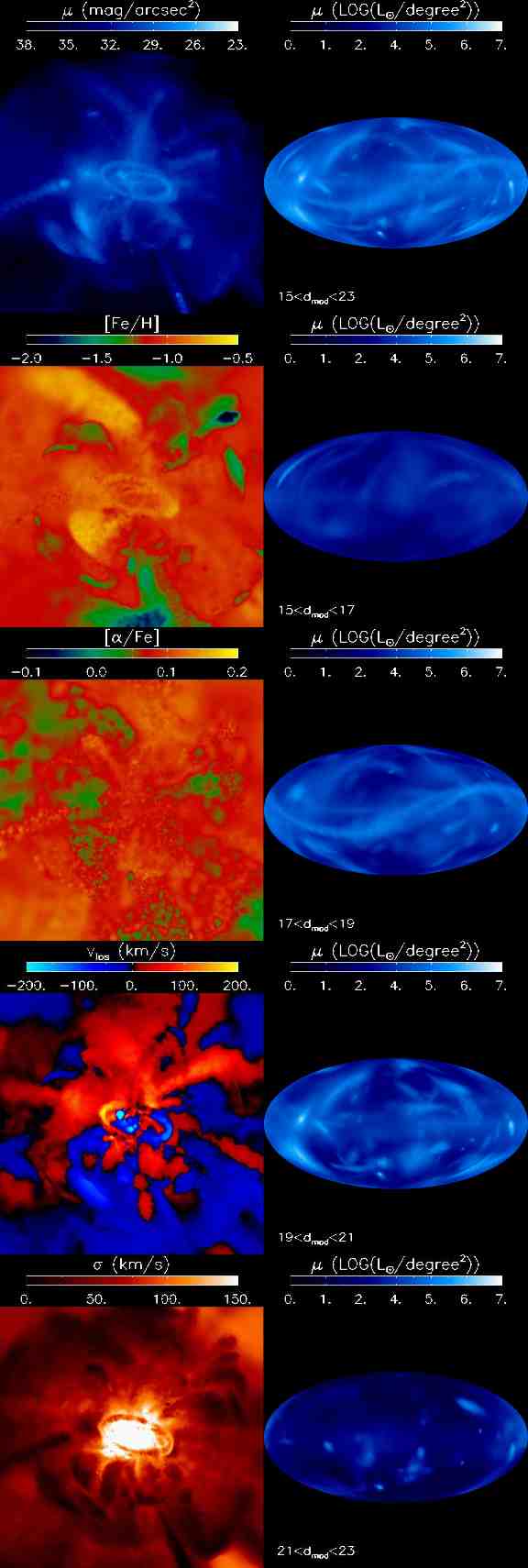}
\caption{\label{oldyoung.fig}
As Figure \ref{halo.fig} but for our ``artificial'' halos constructed from the most ancient (left hand panels) and most recent (right hand panels) accretion events.}
\end{figure*}

\begin{figure*}
\plottwo{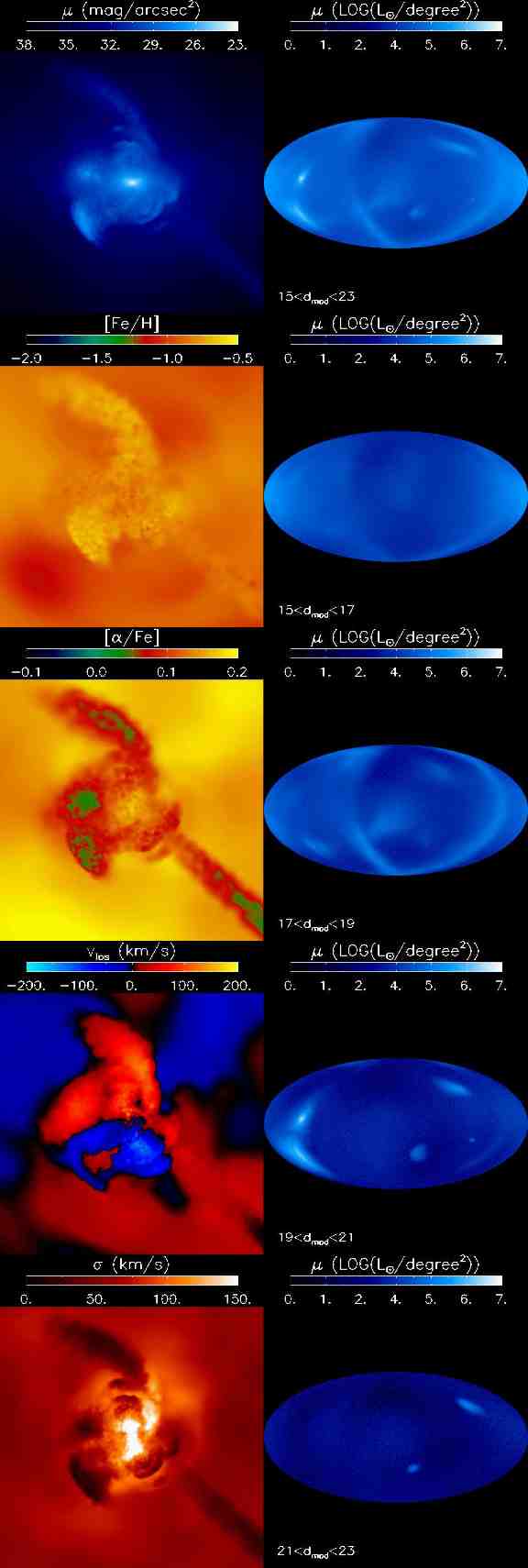}{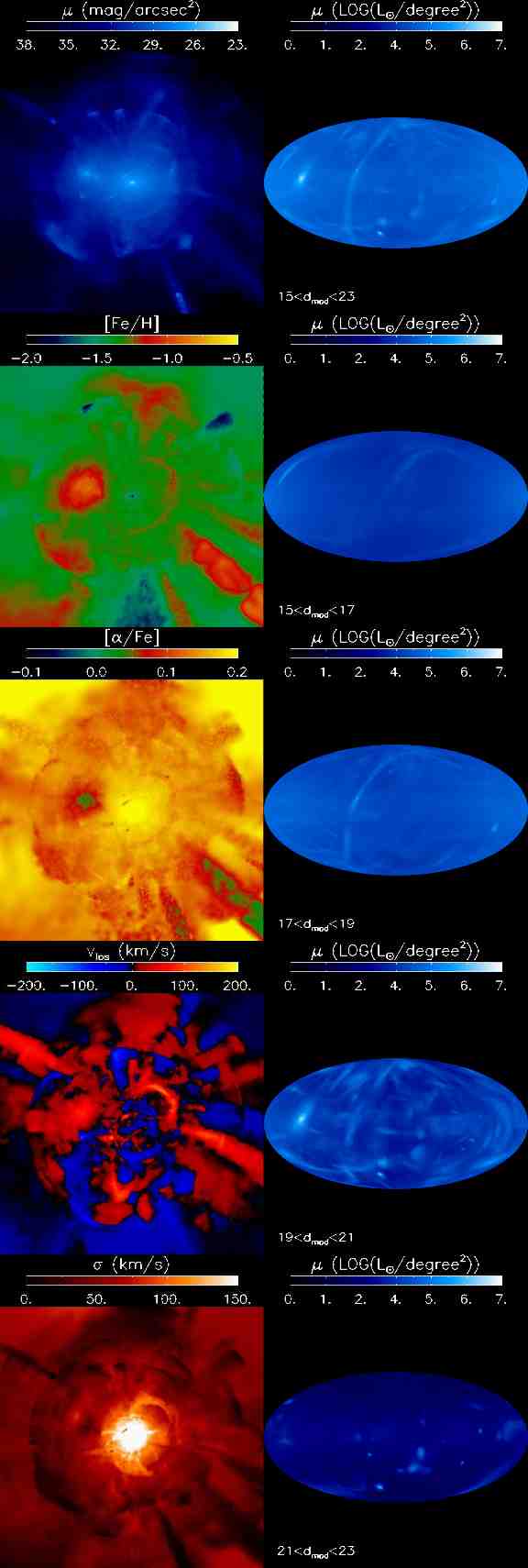}
\caption{\label{highlow.fig}
As Figure \ref{halo.fig} but for our ``artificial'' halo constructed from the highest  (left hand panels) and lowest (right hand panels) luminosity accretion events.}
\end{figure*}

\begin{figure*}
\plottwo{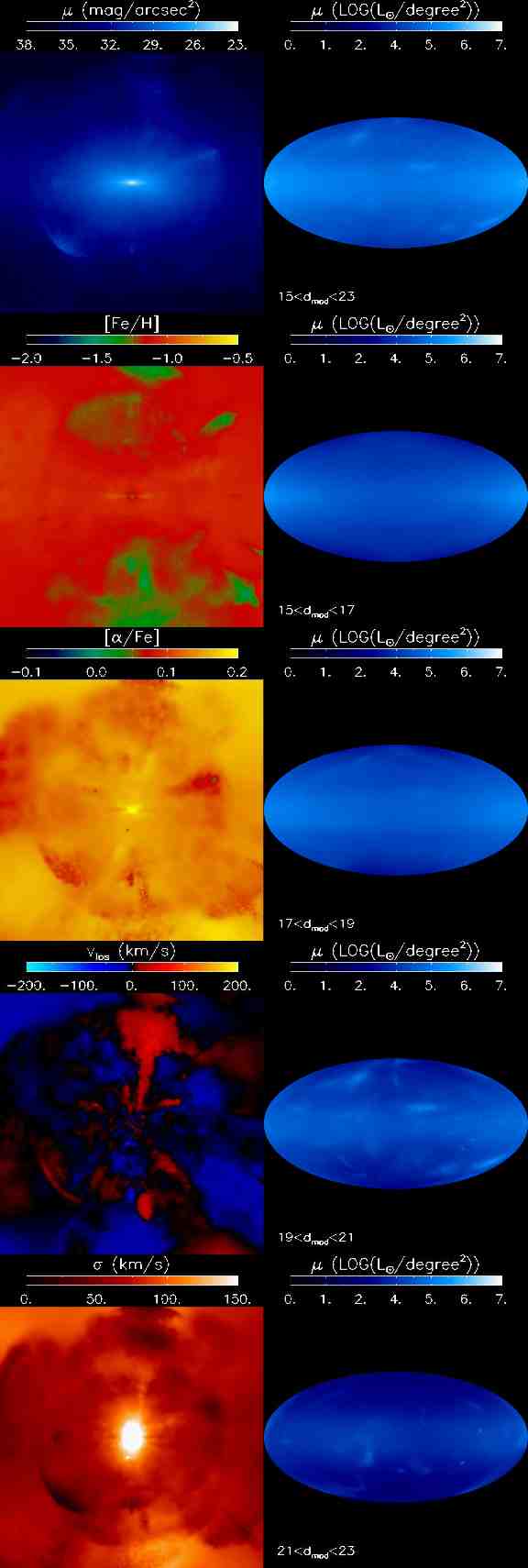}{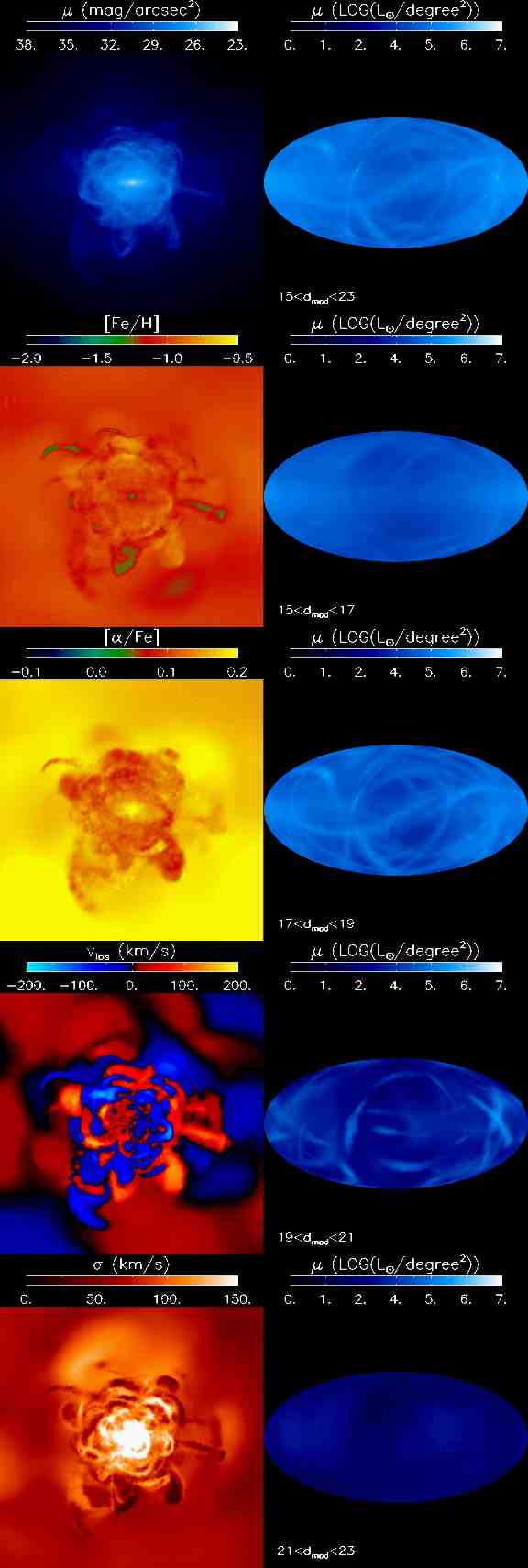}
\caption{\label{radcirc.fig}
As Figure \ref{halo.fig} but for our ``artificial'' halo constructed from the accretion events on the most radial (left hand panels) and most circular (right hand panels) orbits.}
\end{figure*}

The intuition outlined in the preceding paragraph and summarized in Table \ref{trends.tab} is illustrated in Figures \ref{oldyoung.fig},  \ref{highlow.fig} and \ref{radcirc.fig}, which repeat Figure \ref{halo.fig} for our  paired ``ancient'' / ``recent'', ``high luminosity''/``low luminosity'' and ``radial''/``circular'' artificial halos. When compared to our ``standard'' example in Figure \ref{halo.fig} stars in the ``ancient'' halo are concentrated towards its center with no obvious substructure, small changes in line-of-sight velocity across the halo, large velocity dispersion in its outskirts, slightly depressed [Fe/H] and enhanced [$\alpha$/Fe]. In contrast, stars in the ``recent'' halo are almost entirely in substructure with no obvious smooth component, larger extremes in line-of-sight velocity across the halo, small velocity dispersion in its outskirts, slightly enhanced [Fe/H] and significantly depressed [$\alpha$/Fe]. Both the ``high luminosity'' and ``low luminosity'' halos show some smooth component, but the former contains a few broad, bright substructures while the latter is crossed by many thin substructures. In the ``high luminosity'' halo [Fe/H]  is significantly enhanced, while it is depressed in the ``low luminosity'' case. Finally, stars in the ``radial'' halo are smoothly distributed with some substructure in the form of shells, plumes or clouds while stars in the ``circular'' halo show abundant substructure in the form of rosettes or great circles.

\subsection{Worked examples}

\begin{figure*}
\plottwo{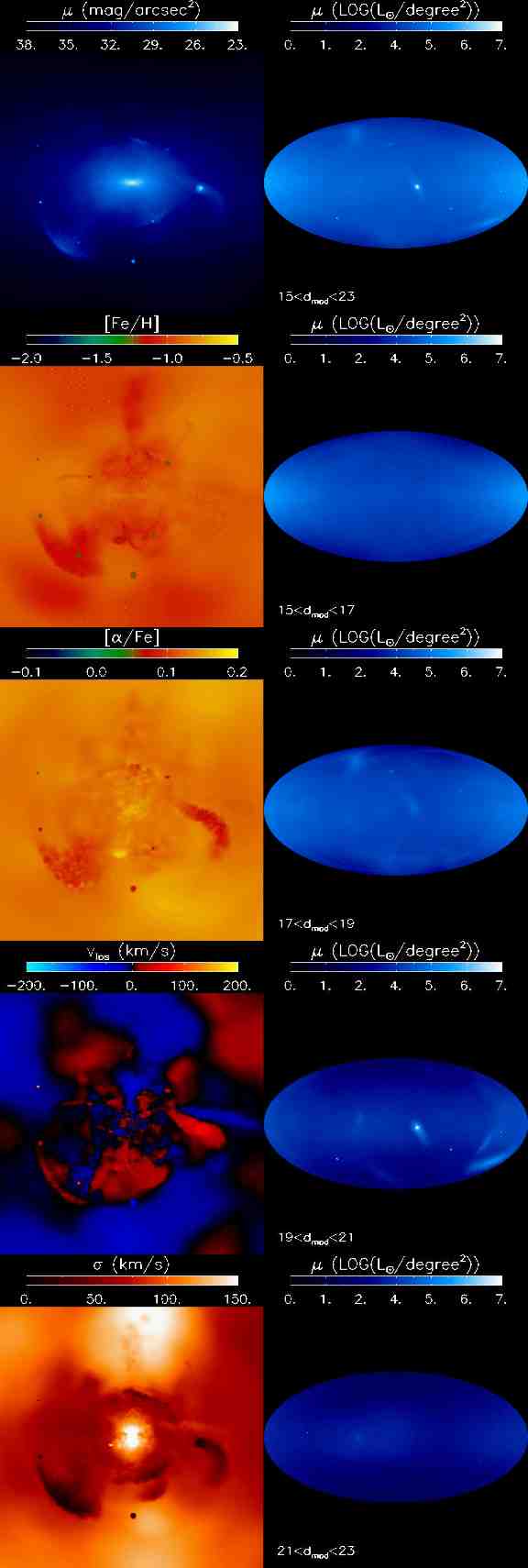}{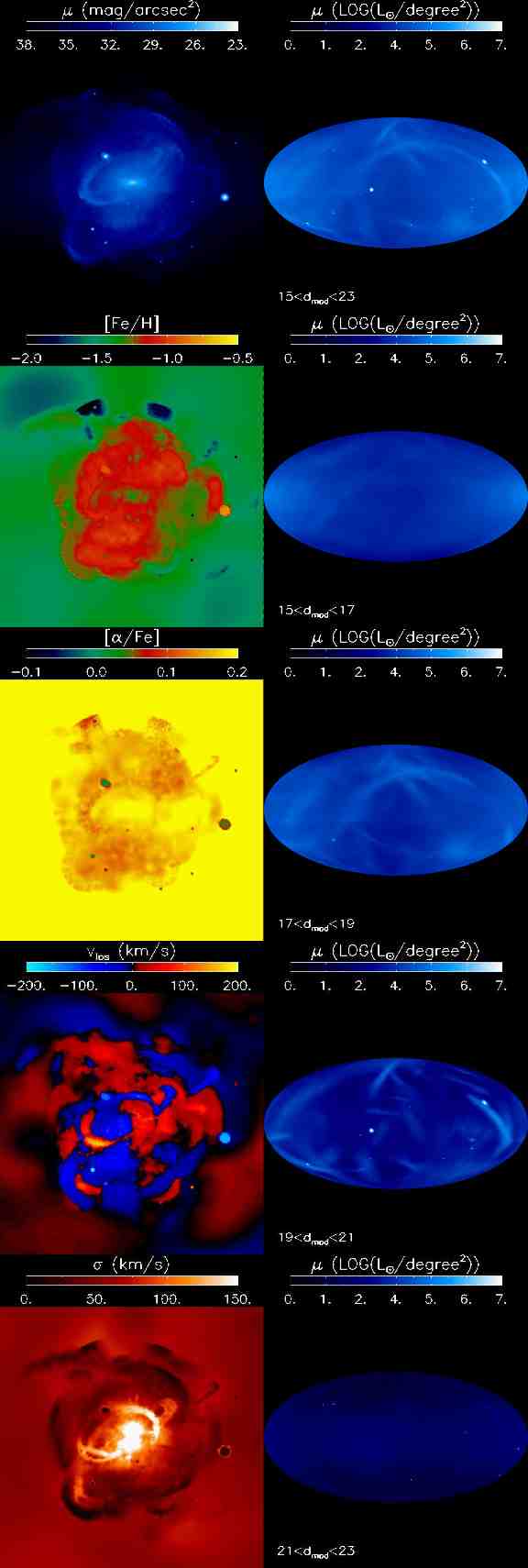}
\caption{\label{halo_9_10.fig}
Repeats Figure \ref{halo.fig} for ``standard'' halos 4 and 6.}
\end{figure*}

Our ``artificial'' halos were deliberately constructed to be outliers through the manipulation of their accretion histories (as shown by the colored lines in Figure \ref{buildup.fig}). Figure \ref{halo_9_10.fig} shows two clear examples of outliers in ``observations'' of our  ``standard'' halo models (halos 4 and 6), which we can use, along with halo 8 (Figure \ref{halo.fig}) and Table \ref{trends.tab}, to test to whether our intuitive interpretations are accurate.

Consider first the spatial and kinematic distributions in Figure \ref{halo.fig} (halo 8).
There are several high surface brightness features on a variety of spatial scales that indicate recent accretion of both low and high-mass objects. These recent events have not yet fully mixed, so there are also large areas of low-dispersion  in the outskirts of the galaxy, with the average line-of-sight velocity varying between extremes of $\pm$ 150 km/s. The mixture of cloud and great circle morphologies suggest a range of orbital properties in these recent events.

These interpretations are reinforced by the
spatial distribution of the abundance patterns.
In [Fe/H] halo 8 shows an average background of $\sim -1.0$, with clear distinct substructures superposed and offset by $\sim$0.5 dex to both higher and lower metallicity. 
These substructures are uniformly lower in [$\alpha$/Fe] ($\sim 0.05-0.1$) than the background ($\sim 0.2$). 
The distinct morphology and low [$\alpha$/Fe] of these substructures suggest recent ($< 8$Gyrs ago) accretion, with their variety of metallicities pointing to both low- and high-luminosity progenitors.
The background, fully mixed debris probes further back in the galaxy's history than the distinct substructure.  The high [$\alpha$/Fe] and intermediate [Fe/H] values of this material are indicative of an epoch of early accretion ($\sim$ 10 Gyrs ago) of intermediate mass objects.

In contrast to halo 8, the surface brightness distribution of halo 4 (left hand panels of Figure \ref{halo_9_10.fig}) is relatively smooth, with only moderate ($\pm$ 50 km/s) variations in line-of-sight velocity and high dispersion in the outskirts of the galaxy, suggesting a well-mixed population with little recent accretion. The lack of great-circle morphologies points towards a lack of recent accretions on circular orbits.
The background of halo 4 has a higher average metallicity and slightly lower [$\alpha$/Fe] than halo 8. The high background [Fe/H] suggests significant contribution from just a few high-mass progenitors, while the low [$\alpha$/Fe] suggests that the main accretion epoch was slightly later than usual. 

Finally, the surface brightness plots for halo 6 (right-hand panels of Figure \ref{halo_9_10.fig})  show many small-scale substructures that are not yet fully-mixed, and tend to have great-circle morphologies. 
The outskirts of halo 6 has low [Fe/H] and high [$\alpha$/Fe], while the inner parts have intermediate [Fe/H] and [$\alpha$/Fe]. The former should come from  a large number of low-luminosity, early accretion events, while the latter indicates some intermediate mass and age events. 


\begin{figure}[t]
\plotone{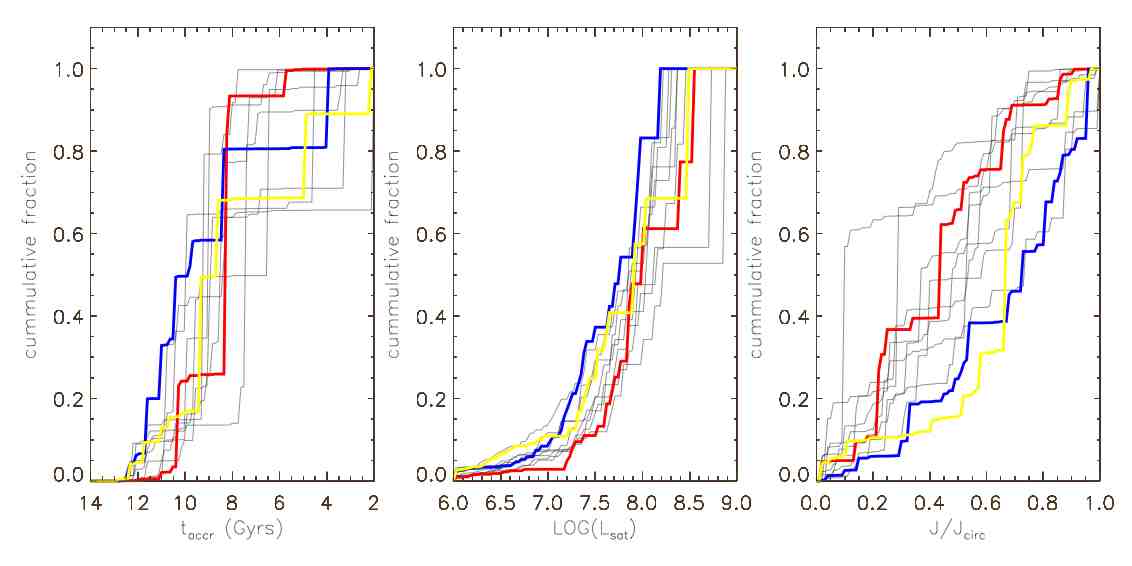}
\caption{\label{buildeg.fig}
Repeats Figure \ref{buildup.fig} for our standard halos, with the histories for halos 4, 6 and 8 highlighted in red, blue and yellow respectively.}
\end{figure}


Figure \ref{buildeg.fig}, which repeats Figure \ref{buildup.fig} with halos 4, 6 and 8 highlighted in color confirms our general conclusions: 
$\sim$70\% of halo 8 (shown in yellow) was accreted prior to 8 Gyrs ago, with a the remaining 30\% in few large events since then;
$\sim$60\% of halo 4 (shown in red) comes from a small set of accretion events of intermediate age (all around 8 Gyrs ago) with less than 5\% of the halo accreted since that time and a bias towards radial orbits; and $\sim$60\% of halo 6 (shown in blue) comes from early ($>$10 Gyrs ago) accretions,  with no large ($> 10^8 L_\odot$) events and most on circular orbits.

\section{Discussion: application to observations}
\subsection{Current data}
\label{current.sec}

The Milky Way's stellar halo has been studied in detail locally with solar neighborhood samples \citep[in some cases including full phase- and/or abundance-space information, see e.g.][]{helmi99b} and more deeply (but typically only in small survey patches) using brighter, photometrically selected tracers \citep[e.g. RR Lyraes and giant stars ---][]{ivezic00,vivas01,morrison00,majewski00}. 
Truly global maps  out to distances of nearly 100kpc have been made only in the last few years using carbon stars and M-giants selected from the 2 Micron All-Sky Survey \citep[although limited to a small number of tracers, rough distance estimates and sparse line-of-sight velocity measurements, see][]{ibata01,majewski03}. 
Most recently, the Sloan Digital Sky Survey (SDSS) has provided panoramas of one quarter of the sky as viewed with turnoff star ``glasses'' and out to distances of $\sim 60$ kpc \citep{belokurov06}.
In coordinate space, these studies of our Galaxy have revealed a smooth, metal-poor inner ($<$10kpc) component to its Galactic halo  and several substructures at larger (10-30kpc) radii that as yet have no confirmed progenitor.
Of the substructures, one appears to trace only a mildly eccentric orbit \citep[the Monoceros ring, see][]{newberg02,ibata03}, while there are three examples of diffuse clouds \citep[in the constellations of Virgo, Triangulum-Andromeda and Hercules-Aquila --- see][]{juric08,rochapinto04,majewski04,belokurov07}.
\citet{belokurov06} report the discovery of a great-circle aligned stream \citep[see also][]{grillmair06c,grillmair06d}, which has tentatively been associated with the Ursa Major dwarf galaxy \citep{fellhauer07}, and there are several more such streams from globular clusters \citep{odenkirchen03,grillmair06a,grillmair06b}.
Two of the Milky Way's other satellites \citep[of more than a dozen --- see][]{belokurov06} have material within a few tidal radii of their main bodies with spatial and velocity signatures consistent with  tidal debris, although this interpretation is neither unique nor conclusive \citep[Leo I and Carina, see][]{sohn07,munoz08}.
Only one satellite \citep[the Sagittarius dwarf galaxy, Sgr, see ---][]{ibata94} has definitive evidence for tidal tails. Sgr's debris stretches entirely around the galaxy, is relatively metal-rich and is aligned with a great circle \citep{ibata01,majewski03,law05}. 

Studies of Andromeda have often been limited to look at only a tiny fraction of the stellar halo, for example taking deep color-magnitude diagrams using the Hubble Space Telescope \citep[e.g.][]{ferguson05,brown07}, or selecting giant stars photometrically using narrow-band filters  for subsequent follow-up spectroscopy \citep[e.g.][]{gilbert06}.
The first global views of the halo, first out to distances of $\sim 50$ kpc \citep{ferguson02} and recently expanded to explore one quarter of the galaxy out to $\sim 150$kpc from its center \citep{ibata07} , were put together by plotting overdensities in the number of red stars selected using broad-band filters to have the right colors and magnitudes to be giant stars associated with Andromeda.
These studies have revealed that, like the Milky Way, Andromeda is dominated by substructure beyond its inner 20kpc, most notably by the giant stellar stream \citep{ibata01} which has been mapped to $>$ 100kpc from its center, and whose progenitor was likely a high-mass ($10^9M_\odot$) satellite on a fairly radial orbit \citep{guhathakurta06,font06c,geehan06,fardal06}. The giant stream and some other (possibly associated) substructures within 10's of kpc of Andromeda's center appear to be relatively metal rich \citep[metallicity$\sim -0.7$][]{ferguson05}, while there is increasing evidence for a smoother, more metal-poor extended component at larger radii \citep[metallicity $\sim$-1.3, see][]{irwin05,chapman06,kalirai06}, extending as far as 165kpc from its center \citep{gilbert06}.

In summary,  it is clear that the expectations for galaxies built within a $\Lambda$CDM Universe outlined in \S \ref{summary.sec} are broadly consistent with our knowledge of the Milky Way and Andromeda galaxies.
Comparing specifically, we have found that a stellar halo built entirely from accretion within a  $\Lambda$CDM context should have:
(i) of order 10 \% of its stars is substructure with of a dozen streams brighter than 35th mag arcsec$^{-2}$ (see \S \ref{frequency.sec} and \S  \ref{amount.sec}) as seems to be the case for both Milky Way \citep{bell08} and Andromeda \citep{ibata07};
(ii) the brightest of these streams more metal rich than the smooth component of the halo \citep[see \S \ref{abundances.sec} and][]{font08} as noted for both galaxies by \citet{gilbert08};
(iii) only $\sim$ 1 surviving satellite (Sgr) with extended,  debris streams brighter than 30th mag arcsec$^{-2}$, typically on a mildly eccentric orbit (see \S \ref{survivors.sec});
(iv) debris morphologies consistent with both circular and eccentric orbits --- great-circles or rosettes for the former (e.g. Monoceros, Sgr and the Orphan stream around the Milky Way) and clouds, plumes or shells for the latter (e.g. the Virgo, Triangulum-Andromeda and Hercules-Aquila clouds around the Milky Way and the Giant Stellar Stream around Andromeda).

Finally, note that the question of  whether the smooth components of stellar halos are contributed primarily from stars accreted during earlier events (as we have assumed), or formed {\it in situ} within the primary dark matter halo  \citep[either in an early disk that was subsequently destroyed --- see][--- or an early, gas rich, rapid accretion event]{abadi03}  remains unresolved.

\subsection{Future prospects} 

Current observations do not yet allow us to build global surface density, velocity, metallicity  and abundance maps (i.e. comparable to Figures \ref{halo.fig}-\ref{halo_9_10.fig}) of any halo. However, the summary of data in \S \ref{current.sec} demonstrates that significant portions of both the Milky Way and Andromeda have been surveyed in density and  (in the latter case) in metallicity using  photometric indicators.
Similar data for other galaxies are gradually becoming available \citep[e.g., M33 --][]{mcconnachie06}, with the prospect of moving beyond the Local Group \citep[e.g.][]{seth07} and building samples of dozens of galaxies on the horizon.

In addition, several projects in the next decade promise orders of magnitude increases in the sizes of data sets discussed here: the entire Milky Way halo will be probed by repeated all-sky surveys of stars capable of detecting (for example) RR Lyraes out to hundreds of kpc (e.g. Panoramic Survey Telescope and Rapid Response System and the Large Synoptic Survey Telescope); plans for next generation spectrographs (e.g. the Wide Field Multi-Object Spectrograph) include taking spectra of hundreds of thousands of Milky Way halo stars spread over 10\% of the sky with sufficient resolution to provide estimates of [$\alpha$/Fe] as well as [Fe/H], and with a deep enough magnitude limit to sample the entire halo; the same instrument could map Andromeda's giant stars in position, velocity, [Fe/H] and [$\alpha$/Fe]. 

Coincidentally,   the inner halo ($<$ 10-20kpc) will be dissected in all six phase-space dimensions (e.g. via ESA's GAIA mission), with sufficient numbers of chemical elements measured to track r- and s-process sources as well as [Fe/H] and [$\alpha$/Fe] (i.e. using WFMOS). The potential of such studies has not been explicitly discussed in this paper since the methods for building our simulated halos are least accurate in this regime, and r- and s-process chemical evolution is not included. However, previous work has already pointed the promise of such data sets in phase- \citep{helmi99a,helmi00,knebe05,penarrubia06} and abundance- \citep{blandhawthorn03,fenner06} space and our own results indicate the sensitivity of abundance patterns to accretion history even in 2-dimensions.

Combining these data sets from the Local Group and beyond: (i) the larger number of galaxies in our samples will allow us to move beyond making broad consistency checks (such as outlined in the previous section) to directly constraining the characteristics of and variation among  {\it observed} recent accretion histories  of galaxies; (ii) the increase in the sizes of data sets will increase our sensitivity to the smallest progenitor objects and allow us to examine the entire luminosity function of galactic building blocks; and (iii) the additional dimensions in the data sets (phase- and abundance- space) will provide probes back to the earliest accretion epochs when we expect the bulk of most galaxies were being assembled.

\section{Conclusions}

In this paper we have explored the connection between the merger history of a galaxy and the present day structure of its stellar halo in phase- and abundance-space.
 
We have found that studies of substructure in the surface-density and/or density distribution in stellar halos are sensitive to the recent ($<$ 8 Gyears ago) merging history of a galaxy.
These substructures can tell us about the mass, orbit and accretion time of progenitors, while the proportion of mass in the smooth component reflects the importance of early events (whether in the form of direct accretions from minor or major mergers or  {\it in situ} formation from a heated disk  or monolithic collapse).
For a Milky-Way-type galaxy, built within the context of $\Lambda$CDM, we expect of order few to tens of percent of the halo to be in the form of substructure, for those substructures to be increasingly dominant at large radii and for the inner halo to be relatively smooth.
These expectations are broadly consistent with 
the current data for the Milky Way and Andromeda galaxies .

Our analysis was based on relating the results of a subjective morphological classification of debris morphology from individual accretion events to the properties of the progenitor satellite.
As our sample of galaxies and substructures within galaxies grows, the challenging nature of reaching very low surface brightness, and the multiple dependencies of substructure characteristics mean that interpretations of individual features in terms of progenitor properties may not always be unique. 
Hence, for specific comparisons of data and observations, it makes sense to recast our results in terms of statistical measures of substructure rather than individual interpretations. 
For example, \citet{bell08} report on a preliminary analysis of the SDSS halo turnoff-star data which quantifies the level of substructure via the dispersion in counts around a smooth background.
The same analysis applied to the 
model stellar halos presented in this paper suggests a degree of substructure similar to the observations.
More generally, these statistics need to be designed to be sensitive to differences in accretion history, armed with the knowledge developed here that: (i) the epoch of galaxy formation sets the percentage of the stellar halo contained in phase-space substructure (and its average [$\alpha$/Fe]); (ii) the number and mass scale of recently accreted objects set the number, angular scales (and mean metallicities) of substructure; and (iii) the orbit type of progenitors set the morphology of substructure. 

In contrast to coordinate-space, signatures in abundance-space are not subject to mixing over time so should last indefinitely. This promises a way to look further back time \citep{blandhawthorn03}.
Our own results demonstrate that variations in [$\alpha$/Fe] and [Fe/H] among the smooth components of different stellar halos reflect variations is the dominant epoch of accretion and masses of progenitor objects respectively.
Moving to higher dimensions in abundance space should allow even more precise interpretations.


\acknowledgments
JSB's work was enabled by startup funds at UC Irvine.
KVJ, AF, SS and SNL  were supported through  NSF
CAREER award AST-0133617.
BER gratefully acknowledges the support of a Spitzer Fellowship through a NASA grant administered by the Spitzer Science Center.

The authors thank the Aspen Center for Physics for hosting a Summer workshop in June 2006 which enabled useful discussion in the group, and the referee for helpful comments on the paper.

\appendix
\section{Generating the images from particle data}

The images presented in \S \ref{summary.sec} were created using an adapted version of the multi-dimensional density estimation code EnBiD \citep{sharma06}.  

For the external galaxy images, the smoothing length $h_i$ of an N-body particle with luminosity $m_i$  was first calculated using the $k$ nearest neighbor scheme in 3d space, with $k=64$.  The surface density maps were then generated by calculating the contributions from particles at each pixel point at position ${\bf r}$ in a 512$\times$512 image using the scatter formulation of smooth particle hydrodynamics (SPH) in two dimensions:
\begin{equation}
\rho(r)=\sum_{i} W({\bf r-r}_i;h_i) m_i,
\end{equation}
where $W$ is the kernel function and $m_i$ and $r_i$ are the particle luminosity and position respectively. 
This estimator gives the true density field in the sense that when applied
globally the total mass is conserved. 
The average values of velocities and abundances were estimated using a simple luminosity-weighted averaging:
\begin{equation}
A({\bf r})=\frac{\sum W({\bf r-r}_i;h_i) m_i A_i}{ \sum W({\bf r-r}_i;h_i) m_i } ,    
\end{equation}
where $A$ is the average of quantity $A_i$.

For the Aitoff projections each  N-body particle was split into multiple particles of equal mass and distributed in 3d space using a kernel function and a smoothing length corresponding to 64 neighbors.  The number of particles $n$ that an N-body particle is split into was set equal to the number of pixels the N-body particle encompasses in the Aitoff projection. For an N-body particle of smoothing length $h$ at a distance $r$ , $n=\pi h^2/(S r^2)$,  $S$ being the solid angle of each pixel in the Aitoff image. The particles were then binned in an Aitoff projection to generate luminosity density maps.   The average values of velocities and abundances were calculated (as above) using a luminosity weighted average over the number of particles in each bin.

\end{document}